\documentclass[lettersize,journal]{IEEEtran}
\usepackage{amsmath,amsfonts}
\usepackage{algorithmic}
\usepackage{algorithm}
\usepackage{array}
\usepackage{caption}
\usepackage[caption=false,font=small,labelfont=rm,textfont=rm]{subfig}
\usepackage{textcomp}
\usepackage{geometry}
\usepackage{stfloats}
\usepackage{url}
\usepackage{verbatim}
\usepackage{graphicx}
\usepackage{cite}
\usepackage{titlesec}

\usepackage{amssymb}
\usepackage{color}
\usepackage{setspace}
\usepackage{amsthm}
  
\newcommand*{\QEDB}{\hfill\ensuremath{\square}}
\usepackage{tabularx}
\usepackage{mathrsfs}
\usepackage{multirow}
\usepackage{makecell}
\usepackage{multicol}
\usepackage{float}
\usepackage{color}
\usepackage{booktabs}
\usepackage{colortbl}
\usepackage{mwe}
\usepackage{subfig} 
\usepackage{subfloat}
\DeclareMathOperator{\sgn}{sgn}
\usepackage{bbding}
\usepackage{multirow}
\usepackage{cc4elsart}
\usepackage{nomencl}
\usepackage{ifthen}
\usepackage{chngcntr}
\counterwithout{figure}{section}
\counterwithout{equation}{section}
\counterwithout{table}{subsection}
\counterwithout{table}{section}

\renewcommand{\nomgroup}[1]{%
\ifthenelse{\equal{#1}{C}}{\item[\textit{C. Parameters}]}{%
\ifthenelse{\equal{#1}{D}}{\item[\textit{D. Variables}]}{%
\ifthenelse{\equal{#1}{B}}{\item[\textit{B. Indices and Sets}]}{%
\ifthenelse{\equal{#1}{A}}{\item[\textit{A. Acronyms}]}{}}}}
}
\makenomenclature

\begin{document}
\captionsetup[figure]{name={Fig.},labelsep=period}

\title{Hydrogen Supply Infrastructure Network Planning Approach towards Chicken-egg Conundrum}

\author{Haoran Deng, Bo Yang, Mo-Yuen Chow, Gang Yao, Cailian Chen, and Xinping Guan \vspace{-3em}
\thanks{\it{Corresponding author: Bo.Yang}}
\thanks{The authors are with the Department of Automation, Shanghai Jiao Tong University, Shanghai 200240, China.}}

\markboth{Journal of \LaTeX\ Class Files,~Vol.~14, No.~8, August~2021}%
{Shell \MakeLowercase{\textit{et al.}}: A Sample Article Using IEEEtran.cls for IEEE Journals}


\maketitle

\begin{abstract}
In the early commercialization stage of hydrogen fuel cell vehicles (HFCVs), reasonable hydrogen supply infrastructure (HSI) planning decisions is a premise for promoting the popularization of HFCVs. However, there is a strong causality between HFCVs and hydrogen refueling stations (HRSs): the planning decisions of HRSs could affect the hydrogen refueling demand of HFCVs, and the growth of demand would in turn stimulate the further investment in HRSs, which is also known as the ``chicken and egg'' conundrum. Meanwhile, the hydrogen demand is uncertain with insufficient prior knowledge, and thus there is a decision-dependent uncertainty (DDU) in the planning issue. This poses great challenges to solving the optimization problem. To this end, this work establishes a multi-network HSI planning model coordinating hydrogen, power, and transportation networks. Then, to reflect the causal relationship between HFCVs and HRSs effectively without sufficient historical data, a distributionally robust optimization framework with decision-dependent uncertainty is developed. The uncertainty of hydrogen demand is modeled as a Wasserstein ambiguity set with a decision-dependent empirical probability distribution. Subsequently, to reduce the computational complexity caused by the introduction of a large number of scenarios and high-dimensional nonlinear constraints, we developed an improved distribution shaping method and techniques of scenario and variable reduction to derive the solvable form with less computing burden. Finally, the simulation results demonstrate that this method can reduce costs by at least 10.4\% compared with traditional methods and will be more effective in large-scale HSI planning issues. Further, we put forward effective suggestions for the policymakers and investors to formulate relevant policies and decisions. 
\end{abstract}

\begin{IEEEkeywords}
``Chicken and egg'' conundrum, hydrogen supply infrastructure planning, hydrogen fuel cell vehicle, coordinated multiple networks, decision-dependent uncertainty.
\end{IEEEkeywords}
 \IEEEpeerreviewmaketitle
 \nomenclature[Aa]{HFCV}{Hydrogen fuel cell vehicle.}
\nomenclature[Ac]{HRS}{Hydrogen refueling station.}
\nomenclature[Ad]{RO}{Robust optimization.}
\nomenclature[Ae]{DRO}{Distributionally robust optimization.}
\nomenclature[Af]{SO}{Stochastic optimization.}
\nomenclature[Ag]{DDU}{Decision-dependent uncertainty.}
\nomenclature[Ah]{DDU-DRO}{Distributionally robust optimization problem with a decision-dependent uncertainty.}
\nomenclature[Ai]{HSI}{Hydrogen Supply Infrastructure.}
\nomenclature[Aj]{P2G}{Power to gas.}
\nomenclature[Ak]{PV}{Photovoltaic.}
\nomenclature[Al]{HS}{Hydrogen storage.}
\nomenclature[Am]{OTV}{Other vehicle.}
\nomenclature[Ba]{$i$}{Index of power bus or hydrogen node.}
\nomenclature[Bb]{$l$}{Index of traffic link or hydrogen pipeline.}
\nomenclature[Bc]{$n$, $j$}{Indices of scenarios.}
\nomenclature[Bd]{$t$, $p$, $od$}{Indices of time, traffic path, and O-D pair.}
\nomenclature[Be]{$g$, $\mathcal{G}(i)$}{Index and set of PV.}
\nomenclature[Bf]{$\mathcal{L}_T$, $\mathcal{L}$, $\mathcal{T}$}{Set of traffic links, hydrogen pipelines, and daily 24-hour profiles.}
\nomenclature[Bg]{$\mathcal{B}$}{Set of power buses or hydrogen nodes.}
\nomenclature[Bh]{$\mathcal{L}_s(i)$, $\mathcal{L}_e(i)$}{Sets of pipelines starting and ending with hydrogen node \emph{i}.}
\nomenclature[Bi]{$s(l)$, $e(l)$}{Starting and ending nodes of pipeline \emph{l}.}

\nomenclature[C, 01]{$N_b$, $N_M$, $T$}{Number of nodes, hydrogen source, and time periods (24 hours). }
\nomenclature[C, 02]{$c^{hy,P2G,HS,PIP}$}{Annual investment costs of HRS, P2G, HS, and hydrogen pipeline. }
\nomenclature[C, 03]{$\overline{h}_{i}^{hy,P2G,HS}$, $\underline{h}_{i}^{hy,P2G,HS}$}{Max/min invested capacities of HRS, P2G, HS at node \emph{i}.}
\nomenclature[C, 04]{$Da$}{Total number of days in the planning horizon.}
\nomenclature[C, 05]{$\lambda^{MP,MH}$}{Purchasing costs of electricity and hydrogen.}
\nomenclature[C, 06]{$\lambda^{PVc,TN}$}{Economic coefficients for PV curtailment and traffic congestion time.}
\nomenclature[C, 07]{$\lambda^{epe,hpe}$}{Penalty economic coefficients of unserved electricity load and hydrogen load.}
\nomenclature[C, 08]{$N_{od}$}{Number of paths from \emph{o} to \emph{d}.}
\nomenclature[C, 09]{$C_{l}^{\text{max}}$}{Maximum traffic flow of link \emph{l}.}
\nomenclature[C, 10]{$q_{t}^{od,v}$}{Traffic demand of other vehicle or HFCV.}
\nomenclature[C, 11]{$\delta_{l,p}^{od}$}{Incidence coefficient relating link and path, if path \emph{p} passes link \emph{l}, the value is 1; otherwise, the value is 0.}
\nomenclature[C, 12]{$\delta_{p,i}^{od}$}{Incidence coefficient relating path and node, if path \emph{p} passes node \emph{i} the value is 1; otherwise, the value is 0.}
\nomenclature[C, 14]{$t_{l}^{0}$}{Ideal travel time without traffic congestion.}

\nomenclature[C, 17]{$\eta_{HC}$, $\eta_{HD}$}{Charging and discharging efficiencies of HS.}
\nomenclature[C, 18]{$\overline{F}_{l}$, $\overline{E}_{l}$}{Maximum flow rate and line pack of pipeline \emph{l}.}
\nomenclature[C, 19]{$\overline{H}_{i}^{M,EM}$}{Upper bounds of purchasing hydrogen and electricity quantity from hydrogen source and main grid.}
\nomenclature[C, 20]{$\overline{H}_{i}^{HS}$}{Upper bounds of charging and discharging hydrogen quantity of HS.}
\nomenclature[C, 25]{$b_{line,i}$}{Power flow distribution factor on transmission line due to net injection of bus \emph{i}.}
\nomenclature[C, 26]{$F_{line}^{\text{max}}$}{Capacity of the transmission line.}
\nomenclature[C, 27]{$\underline{pr}$, $\overline{pr}$}{Upper and lower bounds of nodal pressure.}
\nomenclature[C, 28]{$\eta_{P2G}$, $\chi_{P2G}$}{Efficiency of P2G and conversion factor from electricity to hydrogen.}
\nomenclature[C, 29]{$\phi_l$}{Weymouth constant of pipeline $l$.}
\nomenclature[C, 30]{$p_{i,t}^{L}$}{Electrical load.}
\nomenclature[D, 01]{$w_{i}^{hy}$, $w_{i}^{p2g}$, $w_{i}^{HS}$, $w_{l}$}{Binary variables denoting investment status of HRS, P2G, HS, and hydrogen pipeline, 1-invested, 0-not invested.}
\nomenclature[D, 02]{$x_{l,t}$, $t_{l,t}$}{Traffic flow and related travel time of link \emph{l}.}
\nomenclature[D, 04]{$x_{l,t}^{v}$, $f_{p,t}^{od,v}$}{Traffic link and path flow of other vehicles or HFCVs.}
\nomenclature[D, 05]{$x_{i,t}^{HFCV}$}{Traffic flow of HFCV captured by node $i$.}
\nomenclature[D, 07]{$g_{i,t}^{ch}$, $g_{i,t}^{dis}$}{Hydrogen quantities that charge to and discharge from HS.}
\nomenclature[D, 08]{$H_{i,t}$}{Hydrogen quantity stored in HS.}
\nomenclature[D, 09]{$g_{l,t}^{pipin}$, $g_{l,t}^{pipout}$}{Hydrogen inflow to pipeline \emph{l} and outflow from pipeline \emph{l}.}
\nomenclature[D, 10]{$g_{l,t}$}{Hydrogen flow in pipeline \emph{l}.}
\nomenclature[D, 11]{$e_{l,t}^{pip}$}{Line pack of pipeline \emph{l}.}
\nomenclature[D, 12]{$g_{i,t}^{M}$}{Purchasing hydrogen quantity from hydrogen source.}

\nomenclature[D, 16]{$p_{g,t}^{PV}$, $p_{g,t}^{cur}$}{Actual output and the curtailment of PV.}
\nomenclature[D, 17]{$g_{i,t}^{P2G}$}{Hydrogen produced from P2G at node \emph{i}.}

\nomenclature[D, 18]{$g_{i,t}^{sh}$, $p_{i,t}^{sh}$}{Unserved hydrogen and electricity loads.}
\nomenclature[D, 19]{$p_{i,t}^{P2G}$}{Power consumed by P2G at node \emph{i}.}
\nomenclature[D, 20]{$g_{i,t}^{HFCV}$, $g_{i,t}^{D}$}{Hydrogen demand and actual consumed at HRS \emph{i}.}
\nomenclature[D, 21]{$p_{i,t}^{M}$}{Purchasing electricity from the main grid.}
\nomenclature[D, 21]{$pr_{i,t}$}{Nodal pressure at node $i$.}

\printnomenclature[0.85in]


\section{Introduction}\label{sec1}
\IEEEPARstart{A}{s} a clean and zero-carbon fuel, hydrogen has attracted increasing attention for the replacement of fossil fuels to reduce global carbon emissions and improve urban air quality \cite{dodds2015hydrogen,dong2022optimal,liu2021resilient}. Nowadays, traffic exhaust emissions account for 15\% of total carbon emissions. Hydrogen fuel cell vehicles (HFCVs) are regarded as a new type of transportation to replace traditional fuel vehicles in the future. It is believed that HFCVs hold broad application prospects with their high energy density, zero emissions, and rapid refueling characteristics \cite{tao2020collaborative}. It is even expected to replace electric vehicles in the future. Nevertheless, HFCVs have not yet been popularized due to the technical problems of HFCVs and insufficient hydrogen refueling stations (HRSs). Conversely, the low demand for HFCVs results in low utilization of planned HRSs and the waste of resources. Nowadays, the average daily utilization of a single HRS is about 35\%, which is below the level required for economic viability \cite{kurtz2019review}. How to promote the popularization of HFCVs without hydrogen fueling infrastructure, and how to attract investment in HRSs in the absence of HFCVs on the road, is known as the ``chicken and egg'' conundrum. Before HFCVs can be sold, however, at least some HRSs must be operational.

The purpose of this work is to address the difficulty of hydrogen supply infrastructure (HSI) planning caused by the ``chicken and egg'' conundrum in the early stage of HFCV commercialization. There are three critical challenges to overcome. First, infrastructure planning decisions may affect the distribution of traffic flows in the transportation network, while changing the consumption of renewable energy in power network. It is necessary to characterize the collaborative effect among multiple networks. Second, the interaction mechanism of ``chicken and egg'' relationship between vehicles and stations needs to be properly represented. Third, how to construct an investment planning optimization problem framework without sufficient hydrogen refueling demand data is also our concern. The resolution of the aforementioned issues is crucial for the transition from conventional vehicles to HFCVs.

Regarding the first challenge, the coordinated planning problems of HSI have been studied extensively in previous research \cite{wei2021carbon,liu2021optimal}. Authors in \cite{wang2021joint} and \cite{cao2021hydrogen} formulated detailed truck routing, pipeline, and hydrogen storage to quantify the flexibility of the hydrogen transmission systems, and proposed a robust joint planning approach to address various uncertainties from electric load and hydrogen demand. Further, considering the interactions between the power network and transportation network, a two-stage stochastic hydrogen supply infrastructure location planning model was provided in \cite{gan2021multi}. It is worth mentioning that authors in \cite{xiang2021techno} integrated hydrogen and renewable energy into the airport, and studied a novel hydrogen network planning method based on life cycle theory, which improved the integration of the hydrogen system and the airport microgrid system. However, in HSI planning issues, the distance between HRS and the energy supply department (including hydrogen sources and renewable energy generations) directly determines the energy supply cost, and the availability of HRSs relative to HFCVs also affects the convenience and cost of HFCVs refueling. Hence, a coordinated optimization modeling approach that accounts for the inherent contradiction is imperative.

Regarding the second challenge, the availability of HSIs affects the attractiveness of HFCVs for drivers and the hydrogen refueling habits of drivers in the early stage of HFCV commercialization. As the number of HRSs increases, the average distance from the vehicle to the HRS decreases, reducing the time and HFCV mileage required to obtain fuel. It in turn makes HFCVs more attractive, boosting sales and popularity, increasing fuel demand and the profitability of HRSs, leading to further construction of HRSs \cite{keith2020diffusion, kaufmann2021feedbacks}. The above causal relationship between HFCV and, HRS is a typical ``chicken and egg'' conundrum \cite{gis2018hydrogenation}. There are few studies about HSI planning methods considering this critical issue. To this end, a system dynamics model was proposed in \cite{meyer2009modeling} to analyze the feedback effects between HRSs and HFCVs sales. Authors in \cite{ogden2011analysis} developed a station cluster strategy for deploying HRSs. The cluster strategy locates several stations in smaller geographical areas, which creates the potential for station profitability at low levels of HFCVs market penetration \cite{brown2013economic}. Nevertheless, none of these works proposed comprehensive models to characterize this causal relationship in HSI planning, which may result in low utilization of HRSs and wasting of resources.

We can consider this issue from another perspective. The literature \cite{wei2021carbon,liu2021optimal,wang2021joint,cao2021hydrogen,gan2021multi,xiang2021techno} reviewed above all address the uncertainty of hydrogen refueling demand in HSI planning, but they are characterized as exogenous uncertainties that are independent of decisions and resolved automatically over time. However, in the ``chicken and egg'' context mentioned in this work, the larger the scale of HRS investment, the more hydrogen vehicles purchased will be attracted. In other words, the trading market development of HFCVs depends on the siting and sizing planning decisions of HRSs. In turn, hydrogen demand uncertainty for HFCVs needs to be considered in the HRS planning problem. This just forms a decision-dependent relation between hydrogen demand uncertainty and HRS planning decisions. Therefore, it is essential to consider the decision-dependent uncertainty (DDU) of the hydrogen refueling demand of HFCVs in the HRS planning problem.

In real life, some random factor is substantially affected by some decision choices, which is therefore referred to as DDU. For instance, some real production scenarios include investment information affecting production decisions \cite{jonsbraaten1998class}, maintenance decisions affecting component reliability \cite{kobbacy2008complex,zhu2021multicomponent}, and expansion decisions affecting transportation demands. For energy system optimization issues, the prediction error of wind power output and the wind turbine power curve were considered to be dependent on previous investment decisions in \cite{yin2022coordinated} and \cite{yin2019multi}, respectively. It was proved that wind power planning considering DDU can lead to more cost-effective results. In \cite{chen2022robust}, the curtailment strategy was set as influencing the real-time renewable power output, and results demonstrated that the uncertainty level in the power system will be overestimated without taking into account the DDU. Similarly, it is necessary to capture the actual dependent relation existing in HSI planning problem.

Regarding the third challenge, the DDU in \cite{zhang2021robust,zeng2022two,yin2022coordinated,yin2019multi,chen2022robust} is the type that decision-making affects the realization of uncertain factors, which are generally modeled by robust optimization (RO) or stochastic optimization (SO). It is well known that the optimization decision of the RO model is relatively conservative and the application of SO needs to obtain a more accurate probability distribution of uncertainty in advance. However, in the context of the early commercialization of HFCV, the decision maker may not know the exact distribution of such endogenous demand. This probability distribution may be affected by planning decisions. Meanwhile, the support set of hydrogen demand uncertainty will not fluctuate significantly with the change of HRS planning decisions. This just fits into another DDU type: decision-making affects the probability distribution of uncertain factors, whose realization is independent of decisions. This case is generally modeled as a distributionally robust optimization (DRO) problem \cite{deng2023distributionally}, applied to pre-disaster planning, system or road maintenance decisions \cite{luo2020distributionally,noyan2022distributionally,basciftci2020data}. Whereas, there are fewer applications for power network related optimization problems.

In the distributionally robust optimization problem with a decision-dependent uncertainty (DDU-DRO), it is necessary to define an ambiguity set that contains the potential true probability distribution and can reformulate the DRO problem as a tractable deterministic optimization problem \cite{zhou2020linear}. The ambiguity sets in the DDU-DRO model applied include moment-based and Wasserstein metric ambiguity sets. For the moment-based DDU-DRO, the mean and variance of uncertainty factors are decision-dependent, and it was applied in facility location problem of logistics and transportation scenario \cite{basciftci2021distributionally}. However, this ambiguity set cannot guarantee the convergence property of the unknown distribution to the true distribution. For Wasserstein metric DDU-DRO, the empirical distribution is constructed with samples without assigning probability weights. The upper confidence limit is quantified by the Wasserstein radius, it can obtain the results with better out-of-sample performance \cite{mohajerin2018data}. Besides, such an ambiguity set has a potential polyhedral structure, which is more suitable for algorithm development in linear programming and linear conic dual framework \cite{gao2022distributionally}. In the existing literature, there are fewer studies of DDU based on the Wasserstein metric. In \cite{noyan2022distributionally} and \cite{noyan2018distributionally}, the empirical distribution of the Wasserstein metric depends on the decision, but only the case where both the decision variable and the uncertain variable are binary was considered. It is worth mentioning that the decision-making effect on the radius of the Wasserstein ball was modeled in \cite{luo2020distributionally}, but the influence mechanism was not explained. Moreover, in previous studies, it was generally believed that the radius of the ball should be related to the size of samples \cite{zhou2020linear}, and the claim that radius is related to decision-making has no sufficient theoretical basis and needs to be studied deeply. In addition, the method of Wasserstein metric DDU-DRO has not been applied in the power network.

According to the above literature review, there are still some difficulties in the study of HSI planning. First, in terms of modeling, the HRS siting planning decision could affect both the costs of investment and HFCVs traveling. The convenience of HFCV hydrogen refueling and the cost of energy supply should be considered simultaneously to balance the contradiction. Meanwhile, the causal relationship between the HRS sizing planning and the uncertain hydrogen refueling demand, which requires appropriate characterization. Second, in the case of insufficient historical data, it is necessary to establish an effective framework for this investment planning problem and propose efficient solving methods. Third, existing investment planning studies only compare the effectiveness and superiority of proposed methods. Whereas, in the early commercialization stage of HFCV, governments and investors urgently need effective guidance and advice on policies and planning decisions to improve the profitability and efficiency.

To fully address the challenges and research gaps mentioned above, we present a unified HSI planning framework for the ``chicken and egg'' conundrum between HRS and HFCV. The main contributions of this work are summarized as follows: 

(1) A rational investment planning model of initial HSI network in the early stage of HFCV commercialization is proposed. Specifically, for HRS siting planning, we establish a coordinated multi-network HSI planning model taking into account the costs of energy supply and HFCVs traveling. For HRS sizing planning, we propose a novel decision-dependent mechanism for infrastructure planning issues to characterize the causal relationship between vehicles and stations properly.

(2) We develop a DDU ambiguity set on the uncertainty of hydrogen refueling demand. The actual probability distribution of hydrogen refueling demand is set to be a Wasserstein ball centered on an empirical probability distribution that depends on the siting planning decision of the HRS. On this basis, we construct a two-stage decision-dependent distributionally robust planning model. The traditional optimization process will be altered due to the introduction of DDU. To this end, we propose an improved distribution shaping method to transform the high-dimensional nonlinear function and develop techniques of scenario and variable reduction to reduce the computational complexity. 

(3) The simulation results indicate the necessity of considering DDU in the infrastructure planning of early stage of HFCVs commercialization, both in terms of economic feasibility and hydrogen production situations. On this basis, we propose excellent suggestions for investors and policymakers.

The rest of the paper is organized as follows. In~\ref{sec2}, the mathematical modeling of the proposed planning method is introduced. In~\ref{sec3}, we construct the decision-dependent DRO formulation. The solution methodology of the DRO problem is derived in~\ref{sec4}. Numerical results and conclusions are analyzed in~\ref{sec5} and~\ref{sec6}, respectively. 

\begin{figure}[htbp]
\centering
\counterwithout{figure}{section}
\includegraphics[width=3.7in]{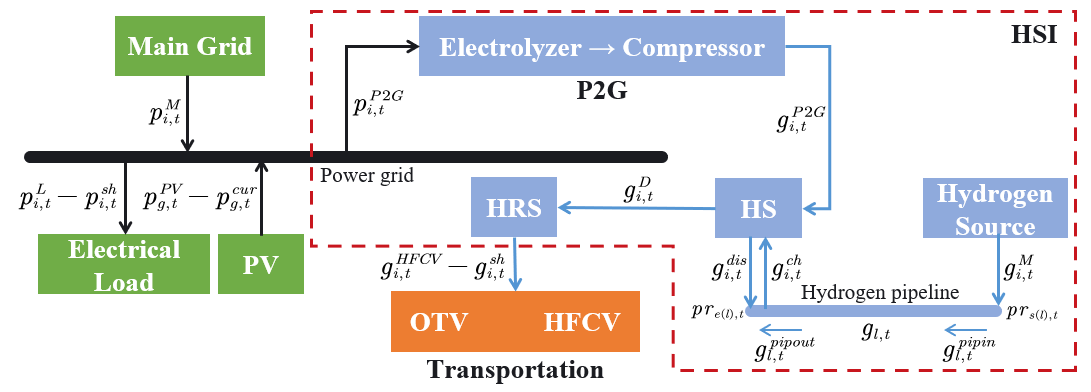}
\caption{Framework of coordinated network.}
\label{fig1}
\end{figure}

\vspace{-0.5cm}

\section{Mathematical Modeling}\label{sec2}
In the proposed hydrogen supply infrastructure (HSI) planning problem, it is empty initially. Various siting and capacity of devices, including HRSs, P2Gs, HSs, and hydrogen pipelines, can be invested to meet energy needs and improve energy efficiency. To fully represent the characteristics of the coupled network, a multi-network collaborative HSI planning model is established in this work. The framework is illustrated in Fig. 1. In this model, the hydrogen network is not only coupled to the power network through the power to gas (P2G) consuming renewable energy but also coordinated with the transportation network through HRS and HFCV. When photovoltaic (PV) output is surplus, the hydrogen network uses redundant PV output through the P2G device and supplies the hydrogen to HFCVs through HRSs. Meanwhile, traffic flows of HFCVs are constrained by the transportation network conditions and influenced by the planning decisions of HRSs. Besides, when the supply of PV is low and cannot meet the hydrogen demand, we introduce hydrogen pipeline planning and allows hydrogen nodes to purchase hydrogen from hydrogen sources. It should be noted that the hydrogen output from HS is not considered to be returned to the power network through fuel cell discharging in this paper, due to the extremely low efficiency of multiple electricity and hydrogen conversions. The fuel cell does not have to be used in the initial infrastructure planning owing to its high operation cost.

\noindent\emph{A. Objective}

To solve the cost contradiction of HFCV and HSI mentioned in~\ref{sec1}, the operation cost of HSIs and the travel cost of HFCVs need to be taken into account simultaneously in the objective function. Therefore, we establish a planning model based on the perspective of the central planner interested in minimizing system costs (or maximizing social welfare). The objective function is formulated as (1), comprising of annual investment cost $C^{inv}$ and the annual system operation cost $C_{t}^{ope}$. The annual investment cost includes that of HRS, P2G, HS, and hydrogen pipelines. The operation cost includes the PV curtailment cost, electricity purchase cost, hydrogen purchase cost, traffic travel time cost of HFCVs, and penalty cost of unserved hydrogen and electricity demands. 
\begin{equation}\resizebox{0.35\hsize}{!}{$
\counterwithout{equation}{section}
\min C^{i n v}+D a \sum_t C_t^{o p e}$}
\end{equation}
\begin{equation}\resizebox{0.8\hsize}{!}{$
\begin{aligned}
C^{i n v}=\sum\nolimits_{i} (c^{hy}{h}_i^{hy}+c^{HS}{h}_i^{HS}+c^{P2G}{h}_i^{P 2 G})+\sum\nolimits_{l \in \mathcal{L}} c_l^{P I P} w_l
\end{aligned}$}
\end{equation}
\begin{equation}\resizebox{0.7\hsize}{!}{$
\begin{aligned}
C_t^{o p e}=&\lambda^{P V c} \sum\nolimits_g p_{g, t}^{cur}+\lambda^{M P} \sum\nolimits_i p_{i, t}^{M}+\lambda^{M H} \sum\nolimits_i g_{i, t}^{M} \\
&+\lambda^{e p e} \sum\nolimits_i p_{i, t}^{s h}+\lambda^{T N} \sum\nolimits_{l \in \mathcal{L}_T} t_{l, t}^{d e}+\lambda^{h p e} \sum\nolimits_i g_{i, t}^{s h}
\end{aligned}$}
\end{equation}

\noindent\emph{B. Transportation Network Constraints}

In the transportation network, each vehicle has a pair of origin and destination and travels between them, which are named O-D pairs. The traffic demand of O-D pairs is defined as the number of vehicles in each time slot intending to travel from \emph{o} to \emph{d}. There are several paths that can be selected by users between \emph{o} to \emph{d}, and each path passes several links. The traffic flow assignment model is formulated by the following constraints. Constraint (4) denotes that the traffic flow of link \emph{l} is the summation of the traffic flow of HFCVs and other vehicles (OTVs). Also, the link capacity is limited. The traffic demand balance equation is described by (5) and the non-negativity of path flows is limited, $\forall v\in\{HFCV, OTV\}$. Eq. (6) sets that the traffic link flow is equal to the total number of vehicles on the paths through it. Constraint (7) denotes the balanced relationship of the traffic flows of the refueling nodes and paths. Constraint (8) sets that the refueling traffic flow on path $p$ must be less than or equal to the total traffic flows of all links passing the HRS. Constraint (9) denotes that the number of refueling users for one node must be not more than the total traffic flows passed by the node.
\begin{equation}
x_{l,t}=x_{l,t}^{HFCV}+x_{l,t}^{OTV},x_{l,t}\le x_{l}^{\text{max}}(l\in\mathcal{L}_{T})
\end{equation}
\begin{equation}
q_{t}^{od,v}=\sum\nolimits_{p=1}^{N_{od}}f_{p,t}^{od,v},\,\,f_{p,t}^{od,v}\ge0
\end{equation}
\begin{equation}
x_{l,t}^{v}=\sum\nolimits_{od\in \mathcal{R}} \sum\nolimits_{p\in \mathcal{P}_{od}}f_{p,t}^{od,v}\delta_{l,p}^{od}
\end{equation}
\begin{equation}
\sum\nolimits_{i} x_{i, t}^{HFCV}=\sum\nolimits_{o d}q_{ t}^{o d,HFCV}
\end{equation}
\begin{equation}
f_{p, t}^{o d,HFCV} \leq \sum\nolimits_{i} \delta_{p,i}^{o d} x_{i, t}^{HFCV}
\end{equation}
\begin{equation}
x_{i, t}^{HFCV} \leq \sum\nolimits_{o d} \sum\nolimits_{p} {\delta_{p,i}^{o d}}^\top f_{p, t}^{o d,HFCV}
\end{equation}

In this work, the travel time cost is determined by the traffic congestion time, and hydrogen refueling time is not considered due to its short duration. The travel time can be described by a latency function (10), which is also called the Bureau of Public Roads (BPR) function \cite{united1964traffic} and can reflect the delayed travel time in links accurately. In Eq. (10), $C_{l}^{\text{max}}$ is the link flow when $t_{l,t}=1.15t_{l}^{0}$ . The total congestion time of link \emph{l} can be obtained in (11). 
\begin{equation}
t_{l,t}=t_{l}^{0}\begin{bmatrix}
1+0.15(x_{l,t}/C_{l}^{\text{max}})^4
\end{bmatrix}\ (l\in \mathcal{L}_{T})
\end{equation}
\begin{equation}
t_{l, t}^{d e}=(t_{l, t}-t_{l}^{0})x_{l,t}=0.15 t_{l}^{0}(x_{l, t})^{5} / (C_{l}^{\text{max}})^{4}
\end{equation}

\noindent\emph{C. Hydrogen Network Constraints}

\emph{1) Infrastructure Investment Constraints}: Constraints (12)-(15) represent the capacity limits of invested devices and the relation among investment decisions. Constraint (16) indicates the node characteristic of radiant hydrogen pipeline network.
\begin{equation}
w_i^{h y} \underline{h}^{hy} \leq h_{i}^{hy} \leq w_i^{h y}\overline{h}^{hy} 
\end{equation}
\begin{equation}
w_i^{HS} \underline{h}^{HS} \leq h_{i}^{HS} \leq w_i^{HS}\overline{h}^{HS} 
\end{equation}
\begin{equation}
w_i^{P2G} \underline{h}^{P2G} \leq h_{i}^{P2G} \leq w_i^{P2G}\overline{h}^{P2G} 
\end{equation}
\begin{equation}
w_i^{P2G}\leq w_i^{hy},  w_i^{hy}= w_i^{HS} (i\in \mathcal{B})
\end{equation}
\begin{equation}
\sum\nolimits_{l \in \mathcal{L}} w_l\leq N_b-N_M
\end{equation}

\emph{2) Hydrogen Demand Constraints}: According to \cite{yao2014multi}, the refueling demand of each node could be approximately calculated as
\begin{equation}
g_{i,t}^{HFCV}=G^{HFCV}f_{t}^{trip}x_{i,t}^{HFCV}/\sum\nolimits_t{f_{t}^{trip}}\sum\nolimits_i{x_{i,t}^{HFCV}}
\end{equation}
where $G^{HFCV}$ is the total hydrogen refueling demand data for HFCVs in this system obtained by investors from government planning departments. $f_{t}^{trip}$ is the trip ratio, which indicates the number of vehicles passing through the traffic network in a certain period of time. Besides, when the HFCVs are not fully popularized, to improve the utilization rate of HRSs, some hydrogen refueling demands can be abandoned appropriately to obtain optimal social benefits. The uncertain refueling demand of each node is the summation of served and unserved hydrogen demand as follows.
\begin{equation}
g_{i,t}^{HFCV}= g_{i,t}^{D}+g_{i,t}^{sh}
\end{equation}
\begin{equation}
g_{i, t}^{sh}\ge 0,\,\,0 \leq g_{i, t}^{D} \leq h_{i}^{hy}
\end{equation}

In constraint (19), the non-negativity of unserved hydrogen demand and the bounds for the served hydrogen demand are ensured. Moreover, to ensure the satisfaction of the energy supply in this system, the actual hydrogen energy provided by HRSs should be higher than a certain percentage of the total hydrogen refueling demand.
\begin{equation}
\sum\nolimits_t{\sum\nolimits_i{g_{i,t}^{D}}}\geqslant \beta G^{HFCV}
\end{equation}
where $\beta$ is the minimum HFCV demand fulfillment rate.

\emph{3) Hydrogen Pipeline Constraints}: According to the gas Weymouth model, the nonlinear relation of hydrogen flow and nodal pressure can be obtained from the discretized equation of motion \cite{osiadacz1987simulation}
\begin{equation}
g_{l,t}^{2}=\phi _l( pr_{s\left( l \right) ,t}^{2}-pr_{e\left( l \right) ,t}^{2} ) \sgn \left( pr_{s\left( l \right) ,t},pr_{e\left( l \right) ,t} \right)  
\end{equation}
\begin{equation}
\underline{pr}\leqslant pr_{i,t}\leqslant \overline{pr} \ \ (l\in \mathcal{L})
\end{equation}
where $\phi_l=\eta_l\pi^2(D_l)^5/16\rho^2Z_lRTL_l f_l$, and $f_l=4( 20.621( D_l ) ^{1/6} ) ^{-2}$ defines the Weymouth friction factor \cite{menon2005gas}. Other parameters can be defined and calculated according to \cite{menon2005gas} and \cite{soave1972equilibrium}. In this work, the gas flow direction of the radial hydrogen pipeline network can be predetermined owing to the known hydrogen source location and HRS candidate nodes. Therefore, the function $\sgn( pr_{s\left( l \right) ,t},pr_{e\left( l \right) ,t} ) $ representing the direction of gas flow in the pipeline is redundant in (21), and the gas flow through a pipeline is positively correlated with the diameter.

In addition, the linepack model is necessary to quantify the amount of gas stored in the pipeline network, which can affect the planning decisions of hydrogen pipelines. In (23), the average gas flow through pipeline $l$ is related to the flows entering and leaving the pipeline. The gas flow rate through the pipeline is limited in (24).
\begin{equation}
g_{l,t}=0.5( g_{l,t}^{pipout}+g_{l,t}^{pipin} ) 
\end{equation}
\begin{equation}
0\leqslant g_{l,t},g_{l,t}^{pipout},g_{l,t}^{pipin}\leqslant w_l\overline{F_l}
\end{equation}

The discretized continuity equations of linepack in pipeline $l$ are
\begin{equation}
e_{l,t}^{pip}=e_{l,t-1}^{pip}+\varDelta t( g_{l,t}^{\mathrm{pipin}}-g_{l,t}^{\mathrm{pipout}} ) 
\end{equation}
\begin{equation}
0 \leq e_{l, t}^{p i p} \leq \overline{E}_l w_l, \,\,e_{l, T}^{p i p}=e_{l, 0}^{p i p}
\end{equation}
where the initial value of the linepack $e_{l, 0}^{p i p}$ can be set as $\varPsi _l\underline{pr}w_l$, and the constant $\varPsi _l=\pi (D_l)^2L_l/4\rho Z_lRT$ \cite{menon2005gas} holds.

\emph{4) HS and Hydrogen Balance Constraints}: The energy balance equation (27) of HS is formulated below \cite{tabandeh2022integrated}. Eq. (28) works as a constraint that limits the maximum amount of hydrogen injection or release of HS. The state of charge (SOC) and the recycle conditions are limited by its capacity in constraint (29).
\begin{equation}
H_{i, t}=H_{i, t-1}+\eta_{H C} g_{i, t{}}^{c h} \Delta t-g_{i, t}^{dis} \Delta t/\eta_{HD}
\end{equation}
\begin{equation}
0 \leq g_{i, t}^{c h},\,\, g_{i, t}^{d i s} \leq \overline{H}_i^{HS} w_i^{HS}
\end{equation}
\begin{equation}
0 \leq H_{i, t} \leq h_i^{HS},\,\, H_{i, 0}=H_{i,T}
\end{equation}

Constraint (30) and (31) denotes the hydrogen balance for each node. Constraint (32) sets the bounds of the hydrogen purchase and the conversion function of P2G.
\begin{small}
\begin{equation}
g_{i, t}^M+g_{i, t}^{P 2 G}+g_{i, t}^{d i s}-g_{i, t}^{c h}=g_{i, t}^{D} 
\end{equation}
\end{small}
\begin{equation}
g_{i, t}^M=\sum\nolimits_{l \in \mathcal{L}_e (i)}g_{l, t}^{pipout}-\sum\nolimits_{l \in \mathcal{L}_s (i)}g_{l, t}^{pipin}
\end{equation}
\begin{equation}
0 \leq g_{i, t}^{M} \leq \overline{H}_{i}^{M}w_{i}^{hy},g_{i, t}^{P 2 G}=\eta_{P 2 G} \chi_{P 2 G} p_{i, t}^{P2G}
\end{equation}

\noindent\emph{D. Power Network Constraints}

The power network constraints include active power balance, the line capacity, and variable constraints, which are referred in Appendix.

\section{Decision-dependent DRO Formulation}\label{sec3}
This section proposes a novel decision-dependent mechanism for infrastructure planning issues to characterize the causal relationship between HFCVs and HRS. For planning the HSIs more reasonably in the early stage of HFCV commercialization without insufficient historical data on hydrogen demand, a DDU Wasserstein ambiguity set for the uncertainty of hydrogen demand is constructed. Then, a DDU-DRO planning model is established.
\vspace{1em}
\noindent\emph{A. Construction of the DDU Ambiguity Set}

\vspace{-1em}For the convenience of expression, the uncertain hydrogen demand is set as $\boldsymbol{u}_t=\left[\boldsymbol{g}_t^{HFCV}\right]_{N_b}$, where $N_b$ is the number of candidate nodes. Assuming that there are $N_s$ possible values of uncertain hydrogen demand at each node, and then the number of scenarios is $N=N_{s}^{N_b}$. The scenarios are indexed by \emph{n}. In DDU-DRO, it is generally assumed that the decision-dependent uncertainty set has finite decision-independent support. In other words, assigning the zero probability to a specific element is equivalent to excluding it from the support, and the mapping from point to set is realized. Meanwhile, the decision dependence is reflected by the varying probability mass function. In this work, the finite support is $\Xi:=\left\{\boldsymbol{u}_t^n\right\}_{n=1}^N$. The candidate probability distributions in $\mathcal{P}\left(\boldsymbol{w}^{h y}\right)$ are represented as a vector $\boldsymbol{p}$, such that $p_{n}(\boldsymbol{w}^{hy})$ is the probability assigned to scenario $\boldsymbol{u}_{t}^{n}$, where $\left\|\boldsymbol{p}\left(\boldsymbol{w}^{h y}\right)\right\|_1=1$. To capture the features of the ambiguity set mentioned in~\ref{sec1}, the Wasserstein metric ambiguity set is expressed as the \emph{Definition 1} in Appendix.

In this work, we consider that the empirical distribution of $\boldsymbol{u}_t$ is related to decision $\boldsymbol{w}^{hy}$, that is, $\boldsymbol{w}^{hy} \rightarrow \hat{p}_n\left(\boldsymbol{w}^{hy}\right) \in \hat{\mathbb{P}}_N$. The probability measure $\hat{\mathbb{P}}_N$ controlling $\boldsymbol{u}_t$ is a function of the decision $\boldsymbol{w}^{hy}$, while the sample space $\Xi$ is independent of $\boldsymbol{w}^{hy}$. It is worth mentioning that different from the construction of Wasserstein metric ambiguity sets proposed by this work, in the majority of DIU optimization problems, the distribution $\hat{\mathbb{P}}_N=\frac{1}{N} \sum_{n=1}^N \delta_{\hat{\boldsymbol{u}}_n}$ is generally adopted as the Wasserstein metric empirical distribution $\hat{\mathbb{P}}_N$ \cite{arrigo2022wasserstein}. In this default empirical distribution, $\hat{\boldsymbol{u}}_n$ belongs to the sample set $\left\{\hat{\boldsymbol{u}}_1, \hat{\boldsymbol{u}}_2, \ldots, \hat{\boldsymbol{u}}_N\right\} $. $\delta_{\hat{\boldsymbol{u}}_n}$ is a Dirac measure on $\hat{\boldsymbol{u}}_n$ and the assumed empirical probability distribution is independent of the location decision. When HFCVs have not yet been popularized, this planning model of HSI will lead to the high conservativeness of decision-making and the waste of resources.

To sum up, different from the \emph{Definition 1} in Appendix, the proposed DDU ambiguity set in this work can be expressed as follows.
\begin{equation}\resizebox{0.89\hsize}{!}{$
\mathcal{P}(\boldsymbol{w}^{hy})=\left\{\boldsymbol{u}^{j}\in \mathbb{P}\Bigg| \begin{array}{l}
\inf \limits_{\Pi} \int_{\Xi^2}\left\|\boldsymbol{u}^j-\boldsymbol{u}^n\right\| \prod\left(d \boldsymbol{u}^j \times d \boldsymbol{u}^n\right) \leq r ,\\ \boldsymbol{u}^n \in \hat{\mathbb{P}}_{N}\left(\boldsymbol{w}^{h y}\right), \forall n, j \in[N], \Pi \in \mathcal{D}(\Xi \times \Xi)
\end{array}\right\}$}
\end{equation}

Since the hydrogen demand considered in this work is with discrete support, the equivalent form of the above ambiguity set is
\begin{equation}\resizebox{0.89\hsize}{!}{$
\mathcal{P}(\boldsymbol{w}^{hy})=\left\{\min\limits _{\boldsymbol{\omega}} \sum_{i,j}\left\|\boldsymbol{u}^j-\boldsymbol{u}^n\right\| \omega_{n j} \leq r \Bigg| \\
\begin{array}{l}
\sum\nolimits_{j=1}^N \omega_{n j}=\hat{p}_n\left(\boldsymbol{w}^{h y}\right),\\
\sum\nolimits_{n=1}^N \omega_{n j}=p_j, \\
\omega_{n j} \geq 0, \forall n, j \in[N]
\end{array}\right\}$}
\end{equation}
where $\boldsymbol{\omega}$ is a joint probability distribution with two marginal distributions given by $\boldsymbol{p}$ and $\hat{\boldsymbol{p}}$,  respectively. $p_j$ is the probability of scenario $\boldsymbol{u}^{j}$. For a given $\boldsymbol{w}^{hy}$, the ambiguity set $\mathcal{P}(\boldsymbol{w}^{hy})$ of the unknown probability distribution depends on the decision variables $\boldsymbol{w}^{hy}$.

\noindent\emph{B. DDU-DRO Formulation}

The original investment planning problem can be equivalent to the following DRO modeling framework.
\begin{equation}
\begin{aligned}
\min _{\boldsymbol{w}}&\{f(\boldsymbol{w})+\sup _{\mathbb{P} \in \mathcal{P}\left(\boldsymbol{w}^{hy}\right)} \mathbb{E}_{\mathbb{\mathbb { P }}}[h(\boldsymbol{w}, \boldsymbol{y}, \boldsymbol{u})]\}\\
\text { s.t. } &\boldsymbol{\varphi}\left(\boldsymbol{w}, \boldsymbol{u}\right) \geq 0 \\
&\boldsymbol{\psi}\left(\boldsymbol{w}, \boldsymbol{y}, \boldsymbol{u}\right) \geq 0
\end{aligned}
\end{equation}

Here $f(\boldsymbol{\cdot})$ and $h(\boldsymbol{\cdot})$ are the investment and operation costs of the system, respectively. $\boldsymbol{w}$ is the vector of HRSs, HSs, and P2Gs investment decisions. $\boldsymbol{y}$ is the vector of operational decision variables. To transform the worst-case expectation of (35) into a tractable form, we develop a lemma giving a reformulation of the above DDU-DRO model for the ambiguity set (34). According to the \emph{Lemma 1} in Appendix, the DDU-DRO problem can be reformulated as follows, and the detailed proof is given in the Appendix.
\begin{small}
\begin{subequations}
\begin{align}
\min _{\boldsymbol{w}}& f\left(\boldsymbol{w}\right)+Da\left[\sum\nolimits_{n=1}^N \hat{p}_n(\boldsymbol{w}^{h y}) \nu_n+r \varepsilon+\eta\right] \\
\text { s.t. } &\nu_n \geq \sum\nolimits_t h(\boldsymbol{w}, \boldsymbol{y}^j, \boldsymbol{u}^j)-\|\boldsymbol{u}^j-\boldsymbol{u}^n\| \varepsilon-\eta \\
&\varepsilon \geq 0\\
&\boldsymbol{\varphi}(\boldsymbol{w}, \boldsymbol{u}^j) \geq 0 \\
&\boldsymbol{\psi}(\boldsymbol{w}, \boldsymbol{y}^j, \boldsymbol{u}^j) \geq 0 \\
&\forall n, j \in[N],\,\, \nu_n,\eta \in \mathbb{R}\notag
\end{align}
\end{subequations}
\end{small}

In this work, the DDU mechanism and complex coordinated network are introduced which lead to the formulated problem containing many nonlinear terms. The scenario probability mapping also needs to be expressed explicitly. Therefore, further reformulation of problem (36) is required, and its detail will be discussed in the next section.

\section{Solution Methodology}\label{sec4}
To obtain a solvable form of the DRO problem in~\ref{sec3}.B and improve solving efficiency, the following transformations are required, including the treatment of the decision-dependent scenario probabilities $\hat{p}_n(\boldsymbol{w}^{h y})$, the linearization of the nonlinear terms, and techniques of scenario and variable reduction.

\noindent\emph{A. Decision-dependent Scenario Probabilities Representation}

In the context of this work, although the realization of the uncertain variable is fixed, its probability depends on the decision. It is very challenging to construct scenario probability mapping $\boldsymbol{x} \rightarrow \hat{p}_n\left(\boldsymbol{x}\right) $ that can properly model the practical problem while maintaining reasonable mathematical tractability.

As mentioned above, the dimension of $\boldsymbol{u}_t$ is $N_b$, and the possible realizations of $\boldsymbol{u}_t$ are $N_s$ discrete values, that is, $\left\{0, 1, \ldots,k,\ldots, N_{s}-1\right\} $. It is assumed that the hydrogen demand in the order of $\left\{0, 1, \ldots, N_{s}-1\right\}$ is monotonically increasing. The parameter vectors $p_0^i=(p_{0,0}^i, p_{0,1}^i, \ldots, p_{0, N_s-1}^i) $$\in[0,1]^{N_s}$ and $p_1^i=\left(p_{1,0}^i, p_{1,1}^i, \ldots, p_{1, N_s-1}^i\right) \in[0,1]^{N_s}$ are obtained from the historical data, which are the probabilities of all possible realizations of the hydrogen refueling demand at node $i$ when $w_{i}^{hy}$ is 0 or 1, respectively. With each sample component $u_{i,t}$, two parameters $p_0^i$ and $p_1^i$ have internal relation $p_{1,0}^{i}\in[0,p_{0,0}^{i}]$ and $p_{1,N_{s}-1}^{i}\in[p_{0,N_{s}-1}^{i},1]$. And then the resulting probabilities as a function of the decision vector $\boldsymbol{w}^{hy}$ is
\begin{equation}
 p_{n, w}^i\left(\left\{u_i=k_i\right\}\right)=(1-w_i^{h y}) p_{0, k_i}^i+w_i^{h y} p_{1, k_i}^i
\end{equation}
where $\sum_{k=0}^{N_s-1} p_{1, k}^i=1$ and $\sum_{k=0}^{N_s-1} p_{0, k}^i=1$. For constructing the empirical probability distribution, the historical data is usually obtained from the hydrogen demand statistics for each node, without considering the specific traffic conditions. Therefore, the empirical probability of each node can be regarded as independent. Under this assumption, the decision-dependent probability distribution $\hat{\mathbb{P}}_{N}(\boldsymbol{w}^{hy})$ is given by the following formula for the probability of scenario \emph{n}:
\begin{equation}\resizebox{0.87\hsize}{!}{$
\hat{p}_n\left(\boldsymbol{w}^{h y}\right)=\prod\limits_{k \in N_s }\prod\limits_{i\in N_b : u_{i, n}=k}\left[(1-w_i^{h y}) p_{0, k}^i+w_i^{h y} p_{1, k}^i\right]$}
\end{equation}

The probability expression (38) is a high-dimensional nonlinear function of the decision variables $w_i^{h y}$, and the number of scenarios is exponential in terms of the number of nodes, which leads to a prohibitively enormous formulation. The construction method proposed is different from that in \cite{noyan2022distributionally} and \cite{noyan2018distributionally}. In this work, the decision variable is binary, and the support of uncertain variables is non-binary discrete. To this end, a key technique called distribution shaping mentioned in \cite{noyan2022distributionally} can be extended according to the context of this paper. This approach can characterize the decision-dependent scenario probabilities via a set of linear constraints.

It can be seen from Eq. (37) that when $\boldsymbol{w}^{h y}$ takes adjacent possible values, the ratio of the probabilities is constant and shown as follows.
\begin{equation}
 \frac{p_{n, w_i^{h y}=1}^i\left(\left\{u_i=k_i\right\}\right)}{p_{n, w_i^{hy}=0}^i\left(\left\{u_i=k_i\right\}\right)}=\frac{p_{1, k_i}^i}{p_{0, k_i}^i} 
\end{equation}

However, to use this property in an optimization model, we can express this scaling for a function of $\boldsymbol{w}^{h y}$. A polyhedral characterization of the resulting distribution can be derived utilizing the feature that $\boldsymbol{w}^{h y}$ is binary. For all $b\in[N_b]$, a truncation vector $\breve{\boldsymbol{w}}_b^{h y}$ is introduced such that it satisfies the following equation.
\begin{equation}
\breve{w}_{b, i}^{h y}=\left\{\begin{array}{ll}
w_i^{h y} & \text { if } 1 \leq i \leq b \\
0 & \text { if }b \leq i \leq N_b
\end{array} \right.
\end{equation}

Then the corresponding scenario probability is expressed as $\pi_{b, n}=\hat{p}_n(\breve{\boldsymbol{w}}_b^{h y}) $. When $\boldsymbol{w}^{h y}=\breve{\boldsymbol{w}}_{N_b}^{h y}$, $\hat{p}_n(\boldsymbol{w}^{h y})=\pi_{N_{b}, n}$ can be deduced for all $n\in[N]$, and the truncation vector $\breve{\boldsymbol{w}}_{b-1}^{h y}$ and $\breve{\boldsymbol{w}}_{b}^{h y}$ have a linear relationship expression. The resulting polyhedral characterization is as following
\begin{equation}\resizebox{0.87\hsize}{!}{$
\begin{cases}\pi_{b, n} \leq \frac{p_{1, k}^b}{p_{0, k}^b} \pi_{b-1, n}+1-w_b^{h y} & \forall k \in\left\{0,1, \ldots, N_s-1\right\}, \\ & \forall b \in\left[N_b\right], \forall n \in[N]: u_b=k \\ \pi_{b, n} \leq \pi_{b-1, n}+w_b^{h y} & \forall b \in\left[N_b\right], \forall n \in[N] \\ \sum\nolimits_{n \in[N]} \pi_{b, n}=1 & \forall b \in\left[N_b\right]\end{cases}$}
\end{equation}
where $\boldsymbol{\pi}\in[0,1]^{N\times N_b}$, $\pi_{0, n}=\prod_k \prod_{i: u_{i, n}=k} p_{0, k}^i$, which represents the baseline probability of scenario \emph{n}. And then the term $\sum_{n=1}^N \hat{p}_n(\boldsymbol{w}^{h y}) \nu_n$ in problem (36) can be reformulated as the mixed integer linear programming (MILP) in (42).
\begin{equation}
\begin{aligned}
\min &\sum\nolimits_{n=1}^N \pi_{N_b, n} \nu_n\\
\text {s.t.} &(41)\\
&\boldsymbol{w}^{h y} \in\{0,1\}^{N_b}
\end{aligned}
\end{equation}

\noindent\emph{B. Linearization of Nonlinear Terms}

We now examine the important case when the norm term is 1-norm distance, that is, the ambiguity set is a Wasserstein-1 ball. Constraint (36b) will take the form
\begin{equation}
\nu_n \geq \sum\nolimits_t h\left(\boldsymbol{w}^{h y}, \boldsymbol{y}^t, \boldsymbol{u}^t\right)-\sum\nolimits_i \gamma_i^{n j} \varepsilon-\eta 
\end{equation}
where the auxiliary variable $\gamma_i^{n j}$ represents the value $|\sum_t u_{i,t}^{n}-\sum_t u_{i,t}^{j}|$. In order to transform (43) into a solvable form, we develop \emph{Lemma 2} in the Appendix, and then constraint (43) can be replaced by the following constraints: 

\begin{small}
\begin{subequations}
\begin{align}
\nu_n &\geq \sum\nolimits_t h(\boldsymbol{w}^{h y}, \boldsymbol{y}^t, \boldsymbol{u}^t)-\sum\nolimits_i z_i^{n j}-\eta \\
z_i^{n j} &\leq \sum\nolimits_t u_{i, t}^n\varepsilon-\sum\nolimits_t u_{i, t}^j\varepsilon+M c_i^{n j} \\
z_i^{n j} &\leq-\sum\nolimits_t u_{i, t}^n\varepsilon+\sum\nolimits_t u_{i, t}^j\varepsilon+M\varepsilon-Mc_i^{n j}\\
z_i^{n j} &\geq( \sum\nolimits_t u_{i, t}^n-\sum\nolimits_t u_{i, t}^j )\varepsilon\\
z_i^{n j} &\geq(-\sum\nolimits_t u_{i, t}^n+\sum\nolimits_t u_{i, t}^j)\varepsilon\\
c_i^{n j} &\geq(\alpha_i^{n j}-1) M_{\varepsilon}+\varepsilon \\
c_i^{n j} &\leq \varepsilon \\
c_i^{n j} &\leq \alpha_i^{n j} M_{\varepsilon}
\end{align}
\end{subequations}
\end{small}

\noindent
where $M_{\varepsilon}$ is the upper bound for $\varepsilon$. The auxiliary vectors are $\boldsymbol{z}\in \mathbb{R}_{+}^{N \times N \times N_b}$, $\boldsymbol{c}\in \mathbb{R}_{+}^{N \times N \times N_b}$, and $\boldsymbol{\alpha} \in\{0,1\}^{N \times N \times N_b}$. Detailed proof is given in the Appendix.

In addition, due to the introduction of the Weymouth model for hydrogen pipelines, there is a nonlinear term in constraint (36e). The relation between average pipeline pressure and node gas pressure is ignored in hydrogen pipeline modeling, and the linepack modeling is simplified. If a detailed linepack model construction is considered here, it can be relaxed to a linear form by referring to \cite{mhanna2021iterative}. Here, we just need to obtain the linear relationship between $g_{l,t}^2$ and $g_{l,t}$ without having to deal with the quadratic term related to gas pressure.

The piecewise linear method \cite{qiu2015linear} is applied to obtain the approximate linearization of (21) as follows.
\begin{small}
\begin{subequations}
\begin{align}
	g_{l,t}&=\sum\nolimits_{k=1}^K{\varDelta g_{l,t}^{\left( k \right)}} \ \ (l\in \mathcal{L})\\
	0 &\leqslant \varDelta g_{l,t}^{\left( k \right)}\leqslant \overline{F_l}/K\\
	g_{l,t}^{2}&=\sum\nolimits_{k=1}^K{\lambda _{G,l}^{\left( k \right)}\varDelta g_{l,t}^{\left( k \right)}}
\end{align}
\end{subequations}
\end{small}

\noindent
where $K$ is the total number of intervals. The length of each interval is $\varDelta g_{l,t}^{\left( k \right)}$. Similar to the above linearization, constraint (11) is linearized as:
\begin{small}
\begin{subequations}
\begin{align}
x_{l,t}&=\sum\nolimits_{h=1}^H{\Delta x_{l,t}^{\left( h \right)}} \ \ (l\in \mathcal{L}_{T})
\\
0&\le \Delta x_{l,t}^{\left( h \right)}\le x_{l}^{\max}/H
\\
t_{l,t}^{de}&=(0.15t_{l}^{0}/\left( C_{l}^{\max} \right) ^4)\sum\nolimits_{h=1}^H{\lambda _{T,l}^{\left( h \right)}\Delta x_{l,t}^{\left( h \right)}}
\end{align}
\end{subequations}
\end{small}

To summarize, the whole model of the DDU-DRO problem (35) is reformulated as an approximate MILP problem \cite{noyan2022distributionally} and thus can be solved efficiently.

\noindent\emph{C. Techniques of Scenario and Variable Reduction}

Although the original investment planning optimization problem can be solved by commercial solvers based on the above reformulation method, the computational complexity is mainly affected by the number of constraints and variables. Constraint (44) involves $(6N^{2} \times N_b +N)$ constraints and $(3N^{2} \times N_b +N+2)$ auxiliary variables, which may pose huge computational challenges as $N$ grows large. Next, effective methods are proposed to reduce the problem size significantly.

\emph{1) Redundant Constraint and Variable Reduction}: In constraint (60) of \emph{Lemma 2}, (60a) and (60b) are equivalent to the inequalities $\gamma_i^{n j}\leq|\sum_t u_{i, t}^n-\sum_t u_{i, t}^j|$. Due to the symmetry of the absolute value function, $\gamma _{i}^{nj}=\gamma _{i}^{jn}$ holds for all $j,n\in \left[ N \right] $. It is sufficient to define only variables $\gamma _{i}^{nj}$ and $\alpha _{i}^{nj}$ that satisfy $n<j$. Therefore, (60c) and (60d) are redundant and can be removed to reduce the number of constraints in this optimization problem.

Next, it is obvious that constraints (60a) and (60b) are Big-M reformulations with binary variables $\alpha _{i}^{nj}$, which lead to huge computational challenges. The norm term in constraint (36b) characterizes the distance of the realizations corresponding to scenarios $n$ and $j$. Then, we assume that the realizations are monotonically increasing \emph{w.r.t.} the indices of the scenarios, that is, the mapping $n\mapsto u_{i,t}^{n}$ or $j\mapsto u_{i,t}^{j}$ is comonotone for all $n,j\in \left[ N \right] $. Therefore, we can further assume that $u_{i,t}^{n}\leq u_{i,t}^{j}$ holds for all $ n<j,\,\,n,j\in \left[ N \right] $, in which equation (47) holds.  In this case, this equation constraint is equivalent to the original absolute value constraint. Further, constraint (60a) and binary variables can be omitted. After the above redundant constraint and variable reduction, constraint (44) involves $(N^{2} \times N_b /2)$ constraints and $(N^{2} \times N_b /2 +2)$ auxiliary variables. The size of the optimization problem is significantly reduced.
\vspace{-0.3em}
\begin{equation}
\gamma_i^{n j}=-\sum\nolimits_t u_{i, t}^n+\sum\nolimits_t u_{i, t}^j 
\end{equation}

\emph{2) Scenario Bundling}: As mentioned above, the number of scenarios for this optimization problem is ${N_s}^{N_b}$, which increases exponentially with the expansion of the investment range. This may pose a huge computational challenge in the MILP problem. We now take focus on the crucial problem of scenario reduction.

We consider grouping scenarios that lead to the same outcome following the scenario bundling technique in \cite{noyan2022distributionally}. An important goal is to identify suitable scenario groups without excessive evaluations of outcome function, to reduce the number of variables in the distribution shaping method. We merge the scenarios with the same outcome into a single expression $\boldsymbol{s}\in \left\{ u_{1,t}^{n},u_{2,t}^{n},* \right\} ^{N_b}$, where the meta-value $*$ denotes changeability. That is, the corresponding outcome is unchanged regardless of its value in the sample set, and this scenario set is denoted as $\mathcal{B} _{\boldsymbol{s}}=\left\{ n\in \left[ N \right] :\,\,u_{i,t}^{n}=s_i\lor s_i=*,\,\,\forall i\in \left[ N_b \right] \right\} $. In the context of our work, if $h( \cdot ) ^n=h( \cdot ) ^j$ holds for all $\forall n,j \in \mathcal{B} _{\boldsymbol{s}}$, we then call $\boldsymbol{s}$ a scenario bundle and denote this common outcome by $h( \cdot ) ^{\boldsymbol{s}}$. Meanwhile, if the elements of a family $S$ of scenario bundles represent a partition $\bigcup_{\boldsymbol{s}\in S}^*{\mathcal{B} _{\boldsymbol{s}}}=\left[ N \right] $ of the scenario set, then we call $S$ a bundling. 

In this work, we assume that the realizations of the uncertain variables are decision-independent. However, in the actual operation of the system, the realization of uncertain hydrogen refueling demand of each candidate node is indirectly limited by constraint (7). Although the support sets of hydrogen demands at HRS candidate nodes are different, there is a portion of considered scenarios not meet the constraints of traffic flow distribution consequentially. We categorize this scenario group into a scenario bundle $\boldsymbol{s}$. For any $*$, it does not meet the actual constraints, and the corresponding outcome is assigned as zero. For the improvement of the distribution shaping method, we define the bundle set $\left\{ \boldsymbol{s}_1,\boldsymbol{s}_2,\cdots ,\boldsymbol{s}_m,\cdots ,\boldsymbol{s}_{N^\prime} \right\} \in S$. The probability measures corresponding to two decision vectors $\left\{ 0,1 \right\} ^{N_b}$ are related as
\begin{equation}
P_1\left( \boldsymbol{s} \right) =P_0\left( \boldsymbol{s} \right) \frac{\sum_{n\in \boldsymbol{s}_m}{p_{1,m}^{n}}}{\sum_{n\in \boldsymbol{s}_m}{p_{0,m}^{n}}}
\end{equation}

Therefore, the polyhedral characterization (41) is reformulated as
\begin{small}
\begin{equation}\resizebox{0.7\hsize}{!}{$
\begin{cases}
	\sigma _{b,\boldsymbol{s}}\le \frac{\sum_{n\in \boldsymbol{s}_m}{p_{1,k}^{b}}}{\sum_{n\in \boldsymbol{s}_m}{p_{0,k}^{b}}}\sigma _{b-1,\boldsymbol{s}}+1-w_{b}^{hy}\\
	\sigma _{b,\boldsymbol{s}}\le \sigma _{b-1,\boldsymbol{s}}+w_{b}^{hy}\\
	\sum_{\boldsymbol{s}}{\sigma _{b,\boldsymbol{s}}}=1\\
\end{cases}$}
\end{equation}
\end{small}

\noindent where the initialization is $\sigma _{0,\boldsymbol{s}}=\prod_{i=1}^{N_b}{\sum_{k\in \boldsymbol{s}_m}{p_{0,k}^{i}}}$. Although we only bundle a portion of special scenarios, it can achieve an order of magnitude reduction on scenarios in practical planning problems. Finally, we can transform the original MILP problem into the following form after reducing the number of scenarios and variables.
\begin{small}
\begin{equation}
\begin{aligned}
\min_{\boldsymbol{w}} &f\left( \boldsymbol{w} \right) +Da( \sum\nolimits_{\boldsymbol{s}}{\sigma _{N_b,\boldsymbol{s}}}\nu _{\boldsymbol{s}}+r\varepsilon +\eta )\\
\text { s.t. } &\nu _{\boldsymbol{s}}\ge \sum\nolimits_t{h}\left( \boldsymbol{w},\boldsymbol{y}^{\boldsymbol{j}},\boldsymbol{u}^{\boldsymbol{j}} \right) -\sum\nolimits_i{\gamma _{i}^{{\boldsymbol{s}}{\boldsymbol{j}}}}\varepsilon -\eta\\
&\left( 4 \right) -\left( 9 \right) ,\left( 12 \right) -\left( 32 \right) ,\left( 36\text{c} \right) \left( 45 \right) -\left( 47 \right) ,\left( 49 \right)\\
&\forall {\boldsymbol{s}},{\boldsymbol{j}}\in [S],\,\,\nu _{\boldsymbol{s}},\eta \in \mathbb{R}
\end{aligned}
\end{equation}
\end{small}

\section{Case Studies}\label{sec5}
In this section, to validate the effectiveness of the proposed model, we apply the proposed HSI planning model to an illustrative system with a 33-bus power network and a 12-node transportation network. This urban transportation network is widely utilized in research related to power and transportation coupled networks \cite{wei2016robust}, which can represent the characteristics of urban traffic effectively. The specific network topologies and system parameters are detailed in the Appendix.

Next, we investigate how the HSI planning decisions and hydrogen production situations are affected by different uncertainty modeling methods in this coupled network. Four cases are considered to demonstrate the effectiveness of the proposed HSI planning model.

Case I (RO): Referring to \cite{gan2020two}, hydrogen demand is set as a robust optimization uncertainty set as follows, and the uncertainty budget is set as 6. 
\begin{equation}\resizebox{0.9\hsize}{!}{$
V=\left\{g_{i, t}^{HFCV} \Bigg| \begin{array}{l}
g_{i, t}^{HFCV}=g_{i, t}^{HFCV,0}+\left(v_{i, t}^{HFCV+}-v_{i, t}^{HFCV-}\right) \Delta g_{i, t}^{HFCV} \\
v_{i, t}^{HFCV+}+v_{i, t}^{HFCV-} \leq 1 \\
\sum\nolimits_{t=1}^T\left(v_{i, t}^{HFCV+}+v_{i, t}^{HFCV-}\right) \leq \Gamma_i^T, \forall t \in \mathcal{T}, i \in \mathcal{B}
\end{array}\right\}$}\notag
\end{equation}

Case II (SO): The SO model is constructed as a sample average model \cite{zhu2019wasserstein}. The samples are generated from the empirical distribution of historical data.

Case III (DRO): The DRO model in reference \cite{arrigo2022wasserstein} is applied to build the ambiguity set of hydrogen demand, in which the empirical probability distribution of the Wasserstein metric ambiguity set is Dirac measure. 

Case IV (DDU-DRO): The DRO model with a decision-dependent ambiguity set on hydrogen demand proposed in this paper is applied. Meanwhile, considering the annual planning of HSI in this paper, three possible realizations of each node are taken to form the support set of hydrogen demand in each time period. The probability values of each possible realization with different planning decisions of each node can be obtained separately according to the principle in~\ref{sec4}.A.

\noindent\emph{A. Planning results comparison}

Fig.2 and Table I show the corresponding devices planning siting, capacity, and costs. In this planning problem, there are three sizes of devices as candidates, including small, medium, and large, respectively.  The total planning cost is the sum of HSI's investment cost and average operation cost. The default hydrogen demand fulfillment rate is set as 0.5. It can be seen that the total investment cost of case I and case II are higher in the four cases. Meanwhile, the total cost of case I is higher than that of case II due to its higher operation cost. In case I, the worst-case scenario for hydrogen demand is considered, resulting in a corresponding increase in operation and investment costs. An SO model is applied in case II with the probability distribution of uncertain parameters closer to the empirical distribution. Although the investment cost is slightly higher than case III and case IV, the operating cost and total cost are the lowest of the four cases. In case III and IV, the ordinary DRO and DDU-DRO models take into account the partial probability information of uncertain parameters, which can effectively reduce the conservativeness of optimization decisions. Therefore, the DRO model can obtain the planning decision with lower investment and operation costs.

\begin{figure}[htbp]
\centering
\includegraphics[width=2.5in]{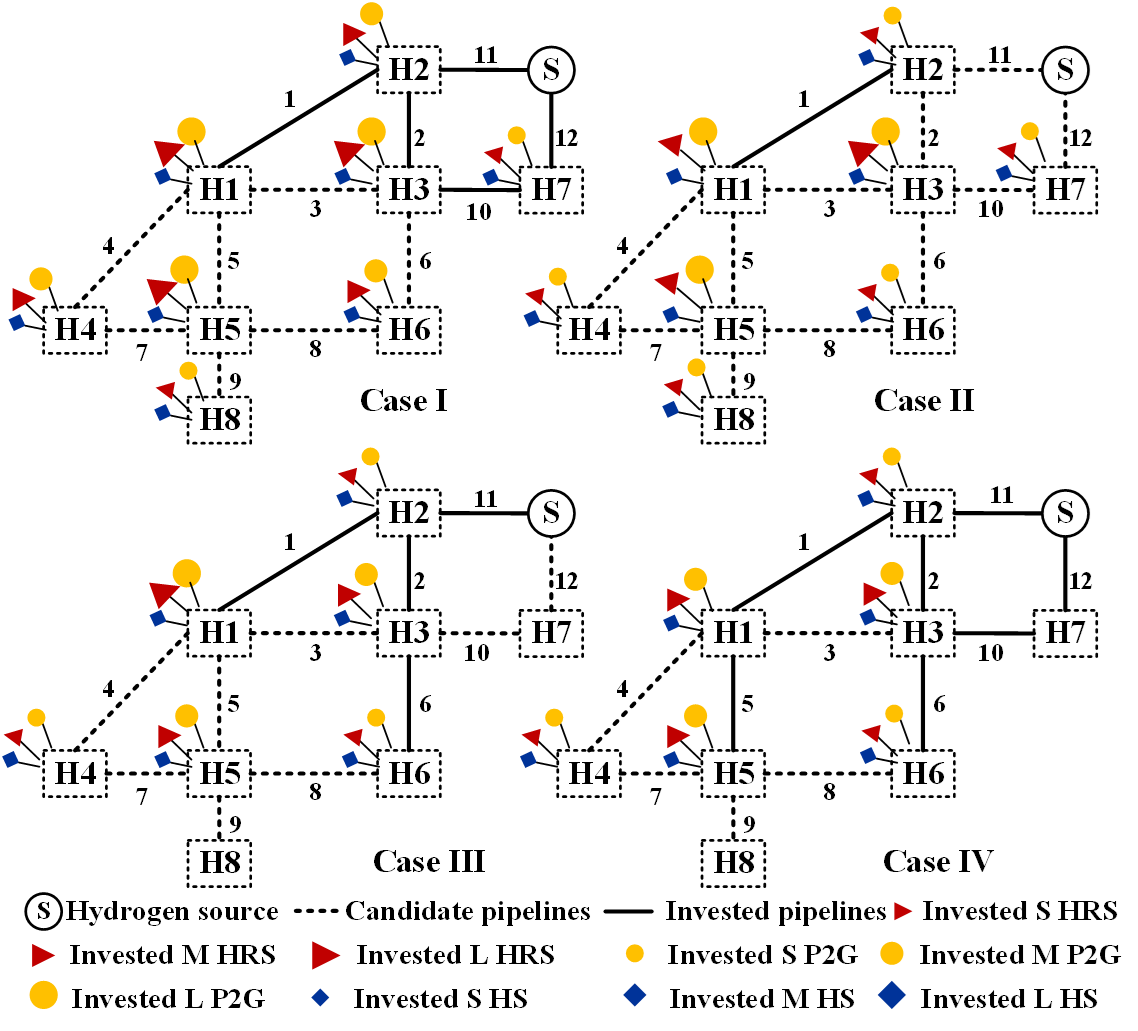}
\caption{HSI devices siting and capacity planning results in case I-IV.}
\label{fig_1}
\end{figure}

\begin{table}[htbp]
\counterwithout{table}{subsection}
\counterwithout{table}{section}
\renewcommand{\thetable}{\Roman{table}}
\renewcommand\arraystretch{1.1}
\caption{Costs allocation in different cases\label{tab:table1}}
\centering
\setlength{\tabcolsep}{0.9mm}{
\scriptsize
\begin{tabular}{|c|c|c|c|c|}
\hline
Costs (M\$)&	Case I	&Case II	&Case III	&Case IV\\
\hline
HRS investment &0.437	&0.363&	0.379	&0.348\\
\hline
P2G investment &11.944	&10.152&	9.068	&7.987\\
\hline
HS investment &0.231	&0.231&	0.202	&0.202\\
\hline
Pipeline investment &0.120	&0.010&	0.100	&0.130\\
\hline
PV curtailment 	&0.000	&0.002	&0.001	&0.004\\
\hline
Purchasing hydrogen &10.898	&0.000	&5.46	&7.260\\
\hline
Traffic congestion time &1.870	&1.250&	4.280	&1.268\\
\hline
Unserved hydrogen demand &1.495	&0.165	&1.739	&1.833\\
\hline
Total cost	&26.995	&12.163&21.230	&19.032\\
\hline
\end{tabular}}
\end{table}

In case IV, the planned devices' capacity, HSI investment cost, and operation cost are all lower than that of case I and III. The total cost of case IV decreases by 10.4\% compared to case III. Specifically, case IV has a reasonable planning for the hydrogen pipeline, with fewer and appropriate planning decisions for P2G. The partial hydrogen demand is satisfied by purchasing hydrogen from the hydrogen source. Case IV applies the DDU-DRO modeling method and considers the influence of planning decisions on the probability distribution of uncertain parameters, so as to appropriately reduce the investment scale and improve the device utilization. Although a small portion of hydrogen demand is unserved in the early stage of commercialization of HFCVs, the best social benefits can be achieved.

\begin{figure}[htbp]
\centering
\includegraphics[width=2.5in]{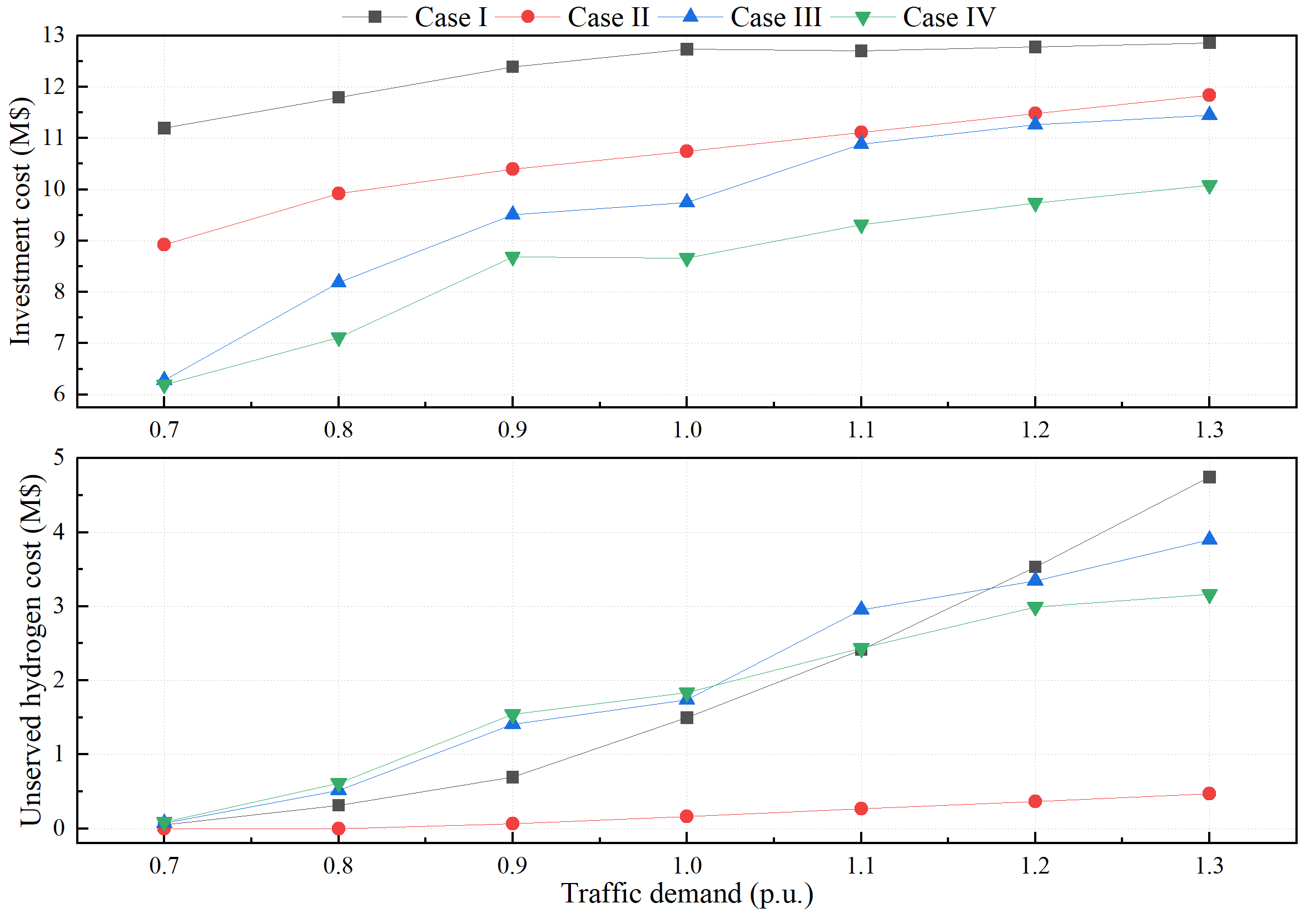}
\caption{Planning results under different traffic demand scales in case I-IV.}
\label{fig_1}
\end{figure}

Next, to illustrate the superiority of the method proposed in this paper under larger traffic demand scales, we compare the solutions of the four cases under different traffic flow scales. Fig. 3 summarizes the performance of each case in terms of HSI investment cost and unserved hydrogen demand cost.

In general, the traffic demand scale can indirectly reflect the hydrogen demand scale of the system. In Fig. 3, the investment cost and the unserved hydrogen cost of the four cases increase with the traffic demand scale, but the investment cost of case IV is always the lowest. When the hydrogen demand scale is smaller, there is a higher unserved hydrogen demand in case IV. However, as the hydrogen demand of the system increases, the unserved hydrogen demand of case IV is relatively smaller in the four cases. This demonstrates the DDU-DRO approach can better adapt to the future increase in HFCVs than the other three approaches. It is worth noting that the unserved hydrogen demand cost in case II is the lowest and fluctuates less with the scale of hydrogen demand. This is because the estimation of uncertain parameters in the SO model is the most accurate and the decision's conservativeness is the lowest, which can achieve lower unserved hydrogen demand to obtain the optimal social benefits.

To sum up, the method proposed in this paper shows significant superiority in economic benefits. It illustrates the necessity of considering DDU in the HSI planning problem in the early stage of commercialization of HFCVs.
	
\noindent\emph{B. Hydrogen supply analysis}

The hydrogen fulfillment rate is an important indicator for hydrogen infrastructure planning, reflecting the ability of the infrastructure to satisfy user demands. In Fig.4, we compare the investment costs and the unserved hydrogen demands of each case under different hydrogen fulfillment ratios. As the hydrogen fulfillment rate increases, the amount of unserved hydrogen demand decreases. But higher investment costs are required to satisfy the high hydrogen fulfillment rate. In addition, when the hydrogen fulfillment rate is higher, the unserved hydrogen demand of case III and IV is approximately similar and lower than that of case I, but case IV achieves this effect with a relatively lower investment cost.

\begin{figure}[htbp]
\centering
\includegraphics[width=2.5in]{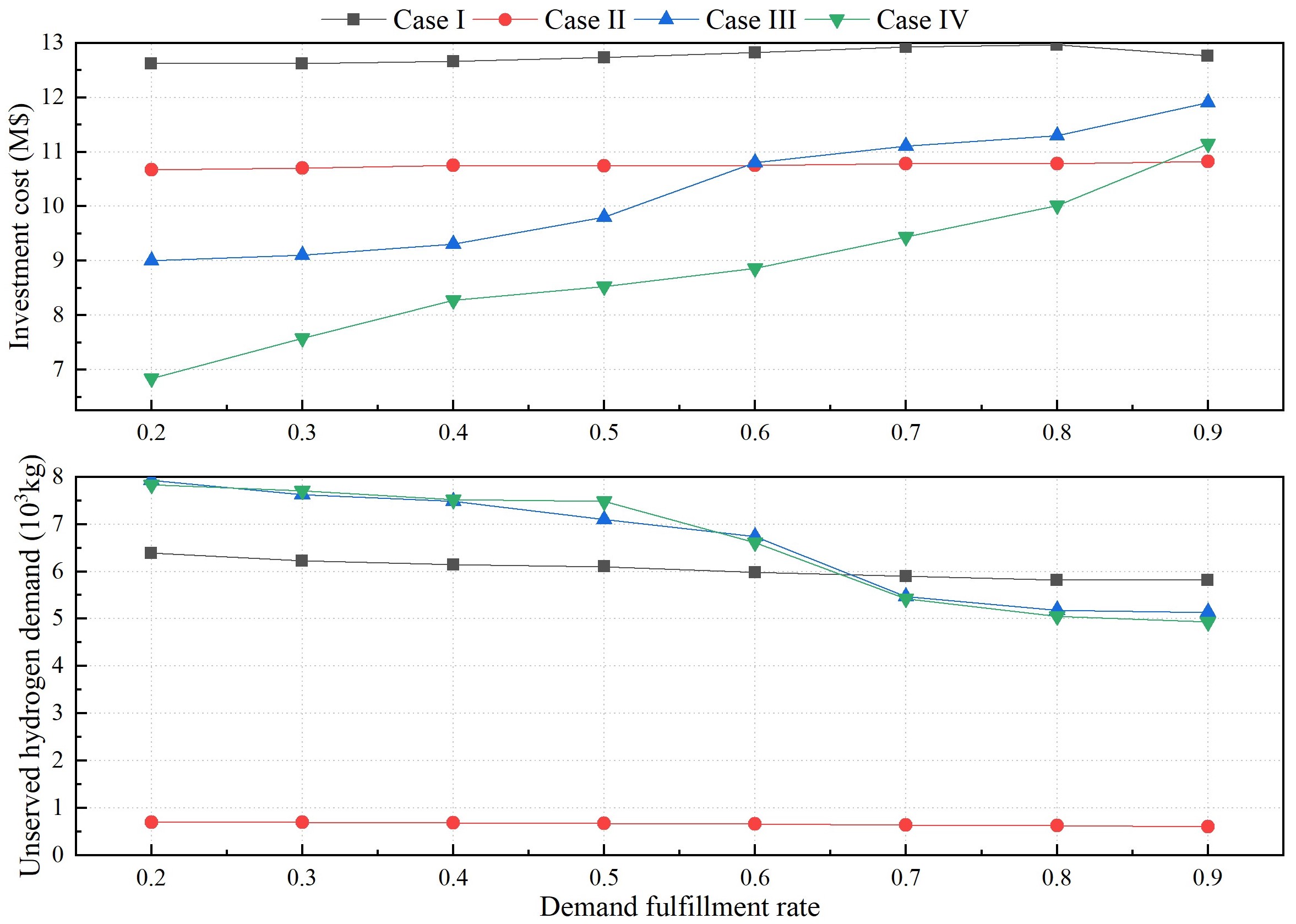}
\caption{Planning results under different hydrogen fulfillment ratios in case I-IV.}
\label{fig_1}
\end{figure}

The results demonstrate that the performance of the proposed method is twofold. On the one hand, in the early stage of infrastructure construction, the government requires a low fulfillment rate generally, which does not exceed 0.5. In this case, the method proposed in this work can satisfy the requirement with a lower investment cost while losing a portion of hydrogen demand. On the other hand, as the reliability requirement for hydrogen infrastructure increases, that is, the fulfillment rate increases, the superiority of the proposed method becomes more obvious. Meanwhile, this also indicates that the method of case IV can better cope with the additional hydrogen demand of the low probability extreme event in the coupled network.

\begin{figure}[htbp]
\centering
\includegraphics[width=2.5in]{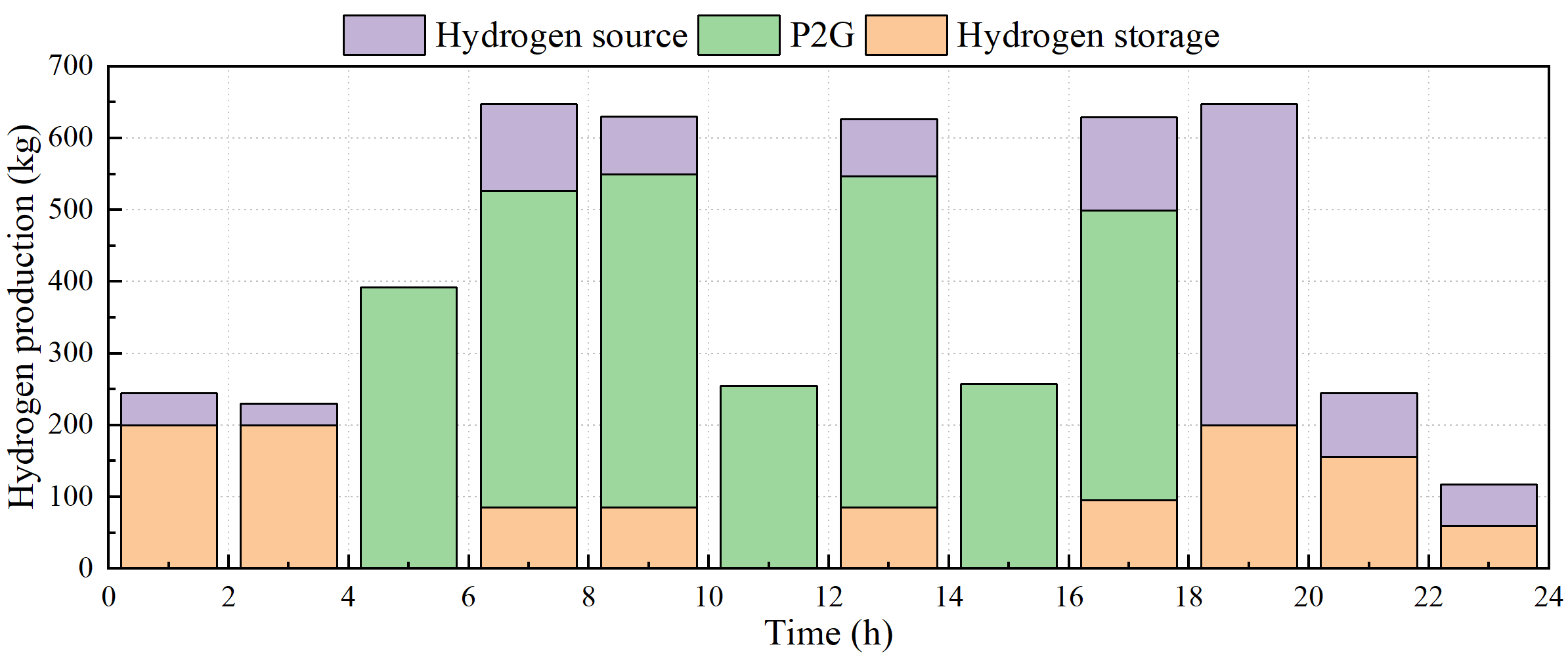}
\caption{Hydrogen production profiles in case IV.}
\label{fig_1}
\end{figure}

In addition, the hydrogen production profiles of case IV during different time periods are shown in Fig. 5. We specifically focus on node 5 for comparison, which is connected to multiple main roads in the transportation network and has a significant hydrogen demand. The results demonstrate that during the peak period from 5:00 to 20:00, the hydrogen demand is primarily met through the P2G conversion of electricity from PV, with any shortfall being supplied by hydrogen source and HS. When PV generation is insufficient from 18:00 to 5:00, hydrogen demand is satisfied only by hydrogen source and HS, which utilizes the excess PV generation conversion to store energy during daylight hours. This indicates that the coupled network is capable of effectively and flexibly utilizing the PV generation to satisfy the hydrogen demand during each time period.

\begin{figure}[htbp]
\centering
\includegraphics[width=2in]{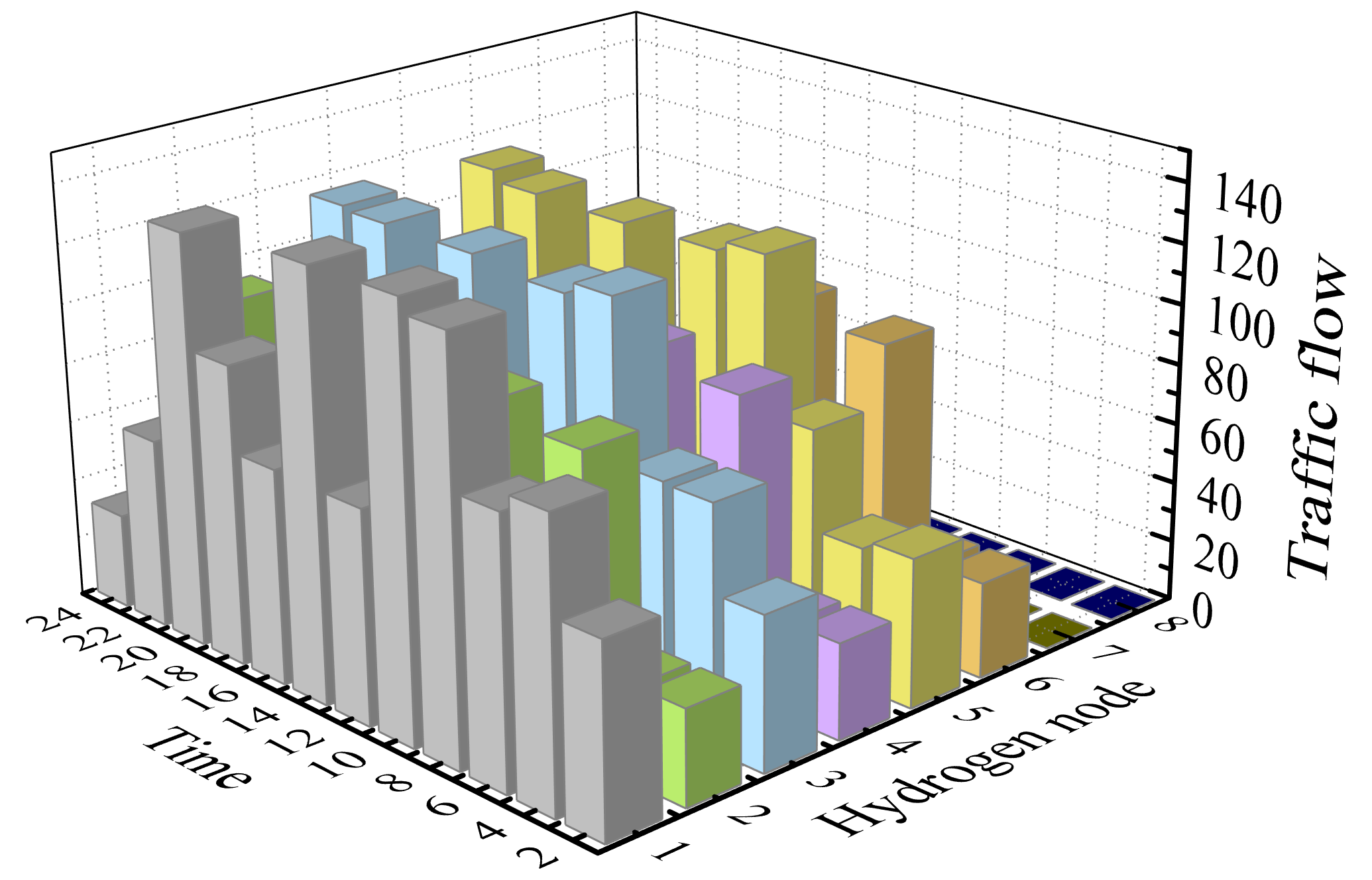}
\caption{Traffic capture characteristics of each node in case IV.}
\label{fig_1}
\end{figure}

Next, the spatial and temporal characteristics of traffic flow capture at each node are shown in Fig. 6. It can be seen that H1, H2, H3, and H5 are the most critical HSI planning candidate nodes capturing a significant portion of traffic flow, and there are all more investments in four cases. Because they are on the way for most paths of all O-D pairs. To fully supply the hydrogen demand of hot spot nodes, case IV forgoes the infrastructure planning for two nodes with relatively low traffic flow. From Table I, it can be observed that hydrogen supply cost and traffic congestion cost of case IV are all at a moderate level among the four cases, balancing the cost contradiction of HFCV and HSI.

\noindent\emph{C. Computational time comparison}

In Table II, we compare the computational time among case III, case IV, and case IV-S, which represents the case IV after scenario and variable reduction. It can be observed that the computational time of case IV is higher than that of case III. However, the approach employed in case IV-S significantly reduces the computational time, even surpassing that of case III. Moreover, it has been verified that both case IV and case IV-S yield the same optimal planning decisions, demonstrating the effectiveness of the proposed method in this work.

\begin{table}[htbp]
\counterwithout{table}{subsection}
\counterwithout{table}{section}
\renewcommand{\thetable}{\Roman{table}}
\renewcommand\arraystretch{1.2}
\caption{Comparison of computational time\label{tab:table1}}
\centering
\setlength{\tabcolsep}{1.5mm}{
\scriptsize
\begin{tabular}{|c|c|c|c|}
\hline
Number of nodes&	Case III	&Case IV	&Case IV-S\\
\hline
4	&769s&	 1152s&	476s\\
\hline
6	&1135s&	1632s&	737s\\
\hline
8	&1399s&	1703s&	804s\\
\hline
\end{tabular}}
\end{table}

\vspace{-1em}

\section{Conclusion}\label{sec6}
This work proposes an HSI planning method to address the "chicken and egg" conundrum between HFCVs and HRSs during the early commercialization stage of HFCVs. We incorporate the distributionally robust characteristic of the uncertain hydrogen demand and establish a mapping relation associated with the empirical distribution of the Wasserstein ambiguity set and the planning decisions of the HRSs. This precisely characterizes the causal relationship between vehicles and stations. Then, we propose an improved distribution shaping method to transform the high-dimensional nonlinear function and develop techniques of scenario and variable reduction to reduce the computational complexity. 

Simulation results demonstrate that this method can obtain more economical and less conservative planning decisions than other models, unnecessary over-planning can be avoided. Meanwhile, this method exhibits greater advantages in the coupled network with larger demand scales and higher hydrogen fulfillment ratio requirements. The effectiveness of the scenario and variable reduction techniques proposed in this work is also verified. 

Further, we propose some suggestions for investors and policymakers:

(1) In the early commercialization stage of HFCV, the method proposed in this work should be applied to make decisions of HSIs planning with demand adaptability and avoid over-planning. This method can fully meet the needs of users while improving energy efficiency.

(2) This method can leave certainly investment space for future HFCV demand growth. Investors can make reasonable additional planning in the long-term planning stage depending on the actual technical development trend and user acceptance of HFCV. The lowest cost investment and the highest equipment utilization rate can be realized.

(3) When the deployment cost of the hydrogen pipeline is relatively low, more hydrogen pipelines can be appropriately invested in the initial HSI network. This can realize a flexible transportation of hydrogen energy among the nodes of the hydrogen network, and lay the foundation for the development of the future hydrogen network.

(4) After the method in this work being applied, it is not necessary for the government to be too conservative in formulating the hydrogen fulfillment ratios, which can be raised appropriately without increasing the cost. This can accelerate the improvement of user satisfaction and promote the popularization of HFCV.

Future work includes investigating the annual co-evolutionary mechanisms between vehicles and stations, as well as researching long-term annual planning methods for hydrogen infrastructure.

\bibliographystyle{IEEEtran}
\bibliography{mybib}

\begin{center}
\textbf{Appendix}\\
\end{center}

\textbf{\emph{Power Network Constraints}}: In this work, DC power flow is used to simulate the operation characteristics of the power network. Although AC power flow is more accurate, it will introduce a large number of nonlinear terms to the optimization problem. In the current context, the power flow parameters such as voltage have no significant impact on the model in this planning problem \cite{mehrjerdi2020wind}. The active power balance and the line capacity are represented in constraints (51) and (52), respectively. Constraint (53) limits the purchasing quantity of electricity from the main grid. The P2G capacity and PV curtailment limit are represented in constraint (54).
\begin{equation}\resizebox{0.7\hsize}{!}{$
\begin{aligned}
\sum_{i \in \mathcal{B}}p_{i, t}^M+\sum_{g \in \mathcal{G}}(p_{g, t}^{P V}-p_{g, t}^{c u r})=\sum_{i \in \mathcal{B}}(p_{i, t}^L+p_{i, t}^{P 2 G}-p_{i, t}^{s h})
\end{aligned}$}\tag {51}
\end{equation}
\begin{equation}\resizebox{1.0\hsize}{!}{$
\begin{aligned}
-F_{\text {line }}^{\max } \leq \sum_{i \in \mathcal{B}} b_{\text {line}, i}[\sum_{g \in \mathcal{G}(i)}\left(p_{g, t}^{P V}-p_{g, t}^{c u r}\right)+p_{i, t}^M-p_{i, t}^L-p_{i, t}^{P 2 G}+p_{i, t}^{s h}] \leq F_{\text {line }}^{\max }
\end{aligned}$}\tag {52}
\end{equation}
\begin{equation}
0 \leq p_{i, t}^M \leq \overline{H}_{i}^{EM}, \,\,p_{i, t}^{s h}\geq 0 \tag {53}
\end{equation}
\begin{equation}
0 \leq p_{i, t}^{P2G} \leq h_{i}^{P2G}, \,\,0 \leq p_{g, t}^{cur} \leq p_{g, t}^{PV} \tag {54}
\end{equation}

\textbf{\emph{Definition 1}}: The Wasserstein metric $ d_W\left(\mathbb{P}_1, \mathbb{P}_2\right)$: $\mathcal{D}(\Xi) \times \mathcal{D}(\Xi) \rightarrow \mathbb{R}$ is defined via \cite{wang2018risk}
\begin{equation}\resizebox{0.87\hsize}{!}{$
d_W\left(\mathbb{P}_1, \mathbb{P}_2\right)=\inf\limits _{\Pi \in \mathcal{D}(\Xi \times \Xi)}\left\{\begin{array}{l}
\int_{\Xi^2}\left\|\boldsymbol{u}_1-\boldsymbol{u}_2\right\| \Pi\left(d \boldsymbol{u}_1 \times d \boldsymbol{u}_2\right): \\
\prod\left(d \boldsymbol{u}_1, \Xi\right)=\mathbb{P}_1\left(d \boldsymbol{u}_1\right), \\
\Pi\left(\Xi, d \boldsymbol{u}_2\right)=\mathbb{P}_2\left(d \boldsymbol{u}_2\right)
\end{array}\right\}$}\tag {55}
\end{equation}
where $\Pi$ is a joint distribution of $\boldsymbol{u}_1$ and $\boldsymbol{u}_2$ with marginal distributions $\mathbb{P}_1$ and $\mathbb{P}_2$, respectively. $\mathcal{D}(\Xi)$ represents the probability distribution set with support $\Xi$, and $\|\boldsymbol{\cdot}\| $ denotes an arbitrary norm. $\|\boldsymbol{u}_1-\boldsymbol{u}_2\| $ is the cost of moving a unit mass from a probability distribution $\mathbb{P}_1$ to $\mathbb{P}_2$. $d_W\left(\mathbb{P}_1, \mathbb{P}_2\right)$ measures the distance between $\mathbb{P}_1$ and $\mathbb{P}_2$.  In this work, the 1-norm ($L_1$-Wasserstein metric) is used due to its superior numerical tractability. Furthermore, the ambiguity set $\mathcal{P}$ is defined as
\begin{equation}\resizebox{0.55\hsize}{!}{$
\mathcal{P}=\left\{\mathbb{P} \in \mathcal{D}(\Xi): d_W\left(\mathbb{P}, \hat{\mathbb{P}}_N\right) \leq r\right\}$}\tag {56}
\end{equation}

This ambiguity set can be regarded as a ball of radius \emph{r} centered at the empirical distribution $ \hat{\mathbb{P}}_N$, and the radius \emph{r} can explicitly control the conservativeness of the decision results.

\textbf{\emph{Lemma 1}}:  If $h(\boldsymbol{\cdot})$ is finite for any $\forall n, j \in[N]$ in (36) and the ambiguity set (34) is nonempty, then the DDU-DRO problem can be reformulated as:
\begin{small}
\begin{subequations}
\begin{align}
\min _{\boldsymbol{w}}& f\left(\boldsymbol{w}\right)+Da\left[\sum_{n=1}^N \hat{p}_n\left(\boldsymbol{w}^{h y}\right) \nu_n+r \varepsilon+\eta\right]\tag {57a} \\
\text { s.t. } &\nu_n \geq \sum_t h\left(\boldsymbol{w}, \boldsymbol{y}^j, \boldsymbol{u}^j\right)-\left\|\boldsymbol{u}^j-\boldsymbol{u}^n\right\| \varepsilon-\eta\tag {57b}\\
&\varepsilon \geq 0\tag {57c}\\
&\boldsymbol{\varphi}\left(\boldsymbol{w}, \boldsymbol{u}^j\right) \geq 0 \tag {57d}\\
&\boldsymbol{\psi}\left(\boldsymbol{w}, \boldsymbol{y}^j, \boldsymbol{u}^j\right) \geq 0 \tag {57e}\\
&\forall n, j \in[N], \,\,\nu_n,\eta \in \mathbb{R}\notag
\end{align}
\end{subequations}
\end{small}

\indent \textbf{\emph{Proof}}: We denote the inner problem $\sup\limits _{\mathbb{P} \in \mathcal{P}\left(\boldsymbol{w}^{hy}\right)} \mathbb{E}_{\mathbb{\mathbb { P }}}\left[h\left(\boldsymbol{w}, \boldsymbol{y}, \boldsymbol{u}\right)\right]$ in (35) as DRO-inner. Since the sample space $\Xi$ is finite, the DRO-inner problem with ambiguity set $\mathcal{P}(\boldsymbol{w}^{hy})$ can be formulated as a linear program as follows.

\begin{small}
\begin{subequations}
\begin{align}
\max _{\boldsymbol{p}, \boldsymbol{\omega}} &\sum_{j=1}^N h\left(\boldsymbol{w}, \boldsymbol{y}^j, \boldsymbol{u}^j\right) p_j \tag {58a} \\
\text { s.t. } &\sum_{j=1}^N \omega_{n j}=\hat{p}_n\left(\boldsymbol{w}^{h y}\right) \tag {58b} \\
&\sum_{n=1}^N \omega_{n j}=p_j \tag {58c} \\
&\sum_{j=1}^N \sum_{n=1}^N\left\|\boldsymbol{u}^j-\boldsymbol{u}^n\right\| \omega_{n j} \leq r \tag {58d} \\
&\sum_{j=1}^N p_j=1\tag {58e} \\
&\omega_{n j} \geq 0, p_j \geq 0, \forall n, j \in[N]\notag
\end{align}
\end{subequations}
\end{small}

After replacing all $p_j$ in (58) with $\sum_{n=1}^N \omega_{n j}$, and then eliminating the redundant constraints (58c) and (58e), the dual of the above linear program is
\begin{equation}\resizebox{0.87\hsize}{!}{$
\begin{aligned}
\min &\sum_{n=1}^N \hat{p}_n\left(\boldsymbol{w}\right) \nu_n+r \varepsilon+\eta\\
\text { s.t. }&\nu_n \geq \sum_t h\left(\boldsymbol{w}^{h y}, \boldsymbol{y}^j, \boldsymbol{u}^j\right)-\left\|\boldsymbol{u}^j-\boldsymbol{u}^n\right\| \varepsilon-\eta\\
&\varepsilon \geq 0, \,\,\nu_n,\eta \in \mathbb{R}
\end{aligned}$}\tag {59} 
\end{equation}
where $\nu_n$ and $\eta$ are auxiliary variables. After substituting (59) into the original DRO model, the desired reformulation (57) can be derived. \QEDB

\textbf{\emph{Lemma 2}}: Constraint (36b) can be replaced by the linear constraint (44).

\indent \textbf{\emph{Proof}}: We introduce auxiliary variables $\gamma_i^{n j}$ and $\alpha_i^{n j}$, and then $\left|\sum_t u_{i, t}^n-\sum_t u_{i, t}^j\right|=\gamma_i^{n j} $ can be linearized by Big-M method as 
\begin{small}
\begin{subequations}
\begin{align}
\gamma_i^{n j} &\leq \sum\limits_t u_{i, t}^n-\sum\limits_t u_{i, t}^j+M \alpha_i^{n j} \tag {60a} \\
\gamma_i^{n j} &\leq-\sum\limits_t u_{i, t}^n+\sum\limits_t u_{i, t}^j+M(1-\alpha_i^{n j})\tag {60b} \\
\gamma_i^{n j} &\geq \sum\limits_t u_{i, t}^n-\sum\limits_t u_{i, t}^j\tag {60c} \\
\gamma_i^{n j} &\geq-\sum\limits_t u_{i, t}^n+\sum\limits_t u_{i, t}^j \tag {60d}\\
\boldsymbol{\alpha} \in&\{0,1\}^{N \times N \times N_b}, \boldsymbol{\gamma} \in \mathbb{R}_{+}^{N \times N \times N_b}\notag
\end{align}
\end{subequations}
\end{small}

Let $\gamma_i^{n j}\varepsilon=z_{i}^{nj}, \boldsymbol{z}\in \mathbb{R}_{+}^{N \times N \times N_b}$, the equivalent form can be obtained as follows.
\begin{small}
\begin{subequations}
\begin{align}
z_i^{n j} &\leq (\sum\limits_t u_{i, t}^n-\sum\limits_t u_{i, t}^j+M \alpha_i^{n j})\varepsilon \tag {61a}\\
z_i^{n j} &\leq[-\sum\limits_t u_{i, t}^n+\sum\limits_t u_{i, t}^j+M(1-\alpha_i^{n j})] \varepsilon\tag {61b}\\
z_i^{n j} &\geq( \sum\limits_t u_{i, t}^n-\sum\limits_t u_{i, t}^j )\varepsilon\tag {61c}\\
z_i^{n j} &\geq(-\sum\limits_t u_{i, t}^n+\sum\limits_t u_{i, t}^j)\varepsilon\tag {61d}
\end{align}
\end{subequations}
\end{small}

There is a bilinear term $\alpha_i^{n j}\varepsilon $ in constraint (61). Auxiliary vector $\boldsymbol{c}\in \mathbb{R}_{+}^{N \times N \times N_b}$ is introduced to relax the non-convex optimization problem into a convex problem, via the set of McCormick inequalities \cite{lv2019optimal}:
\begin{equation}
\left\{\begin{array}{l}
c_i^{n j} \geq \underline{\alpha}_i^{n j} \varepsilon+\alpha_i^{n j} \underline{\varepsilon}-\underline{\alpha}_i^{n j} \underline{\varepsilon} \\
c_i^{n j} \geq \overline{\alpha}_i^{n j} \varepsilon+\alpha_i^{n j} \overline{\varepsilon}-\overline{\alpha}_i^{n j} \overline{\varepsilon} \\
c_i^{n j} \leq \overline{\alpha}_i^{n j} \varepsilon+\alpha_i^{n j} \underline{\varepsilon}-\overline{\alpha}_i^{n j} \underline{\varepsilon} \\
c_i^{n j} \leq \alpha_i^{n j} \overline{\varepsilon}+\underline{\alpha}_i^{n j} \varepsilon-\underline{\alpha}_i^{n j} \overline{\varepsilon}
\end{array}\right.\tag {62}
\end{equation}

Constraint (62) can be simplified to
\begin{equation}
\left\{\begin{array}{l}
c_i^{n j} \geq(\alpha_i^{n j}-1) M_{\varepsilon}+\varepsilon \\
c_i^{n j} \leq \varepsilon \\
c_i^{n j} \leq \alpha_i^{n j} M_{\varepsilon}
\end{array}\right.\tag {63}
\end{equation}
where $M_{\varepsilon}$ is the upper bound for $\varepsilon$. After variable substitution, constraint (36b) can be replaced by constraint (44). \QEDB

~\\

\textbf{\emph{Parameters for Case Studies}}: There are 8 HSI candidate nodes, which are numbered from H1 to H8, and there are 12 candidate hydrogen pipelines among these nodes. Coordinated by the P2Gs and HRSs in HSIs, the topologies of the three networks are shown in Fig.7. There are no prior installations of HSI devices, and the maximum number of HSIs is not required.

The traffic congestion time cost is set as 0.67 \$/min. The penalty prices for unserved electricity and hydrogen demands are set as 130 \$/MW and 245 \$/kg. The unit purchasing prices of electricity and hydrogen from the main grid and hydrogen source are represented in Fig. 8(a). The upper bound for purchasing hydrogen quantity from a hydrogen source is 1000 kg. The upper bound for purchasing electricity quantity from the main grid is 30 MW. The PV curtailment price is set as 40 \$/MW. The investment costs per capacity of HRS, P2G, and HS are 55 \$/kg, 35.1 \$/kW, and 52.5 \$/kg, respectively. Table III provides the major investment parameters for HSI devices \cite{cao2021hydrogen,gan2021multi,xie2020two}. The transportation network data is provided in \cite{gan2020two} and \cite{wei2016robust}. Fig. 8(b) shows the PV generation forecast. The simulations are performed on a laptop with an Intel Core i9-10885H CPU 2.40 GHz using MATLAB with YALMIP and CPLEX 12.9.0 solver.
\begin{table}[htbp]
\counterwithout{table}{subsection}
\counterwithout{table}{section}
\renewcommand{\thetable}{\Roman{3}}
\renewcommand{\thetable}{III}
\renewcommand\arraystretch{1.2}
\caption{Investment parameters for HSI devices\label{tab:table1}}
\centering
\scriptsize
\begin{tabular}{|c|c|}
\hline
Device&Value\\
\hline
HRS capacity (kg) (Max/Min)&5000 / 450\\
\hline
P2G capacity (MW) (Max/Min)&50 / 20\\
\hline
HS capacity (kg) (Max/Min)&6000 / 550\\
\hline
P2G efficiency&0.79\\
\hline
P2G Conversion factor (kg/MW)&28.7\\
\hline
HS efficiency&0.8\\
\hline
Nodal pressure (MPa) (Max/Min)&10 / 3\\
\hline
\end{tabular}
\end{table}

\begin{figure}[htbp]
\addtocounter{figure}{6}
\centering
\counterwithout{figure}{section}
\subfloat[]{\includegraphics[width=2.5in]{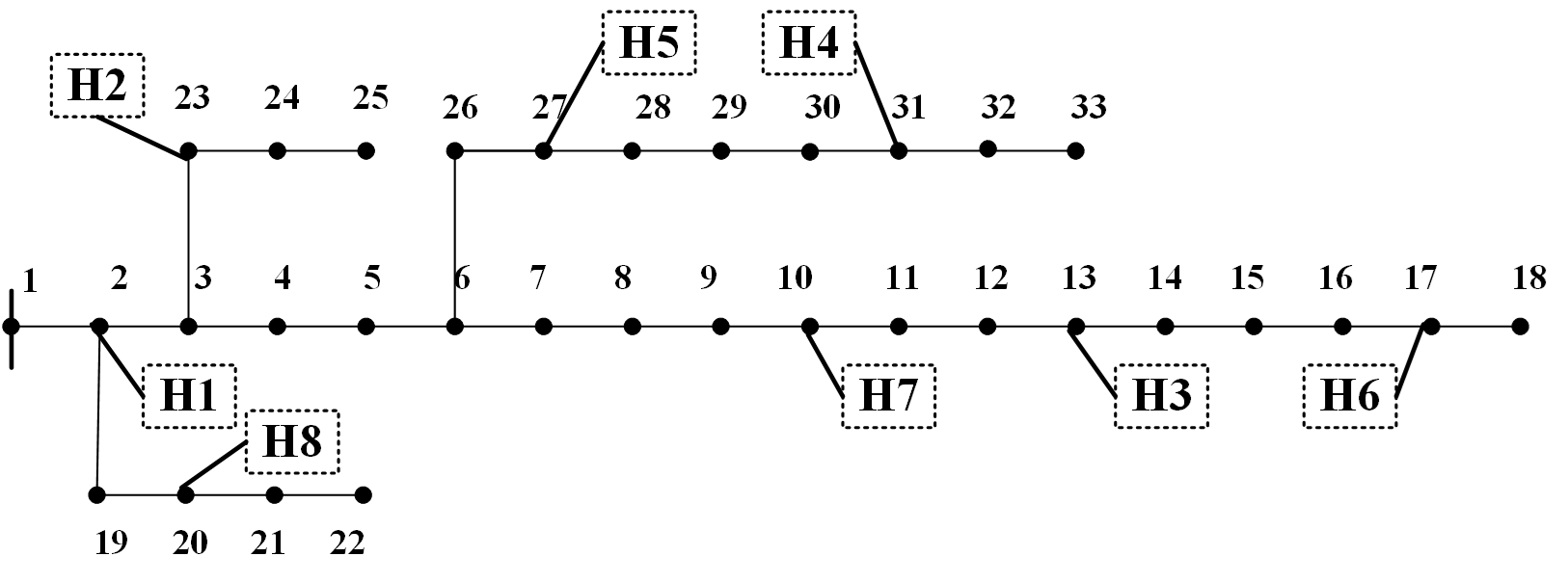}%
\label{fig_first_case}}
\hfil
\subfloat[]{\includegraphics[width=2in]{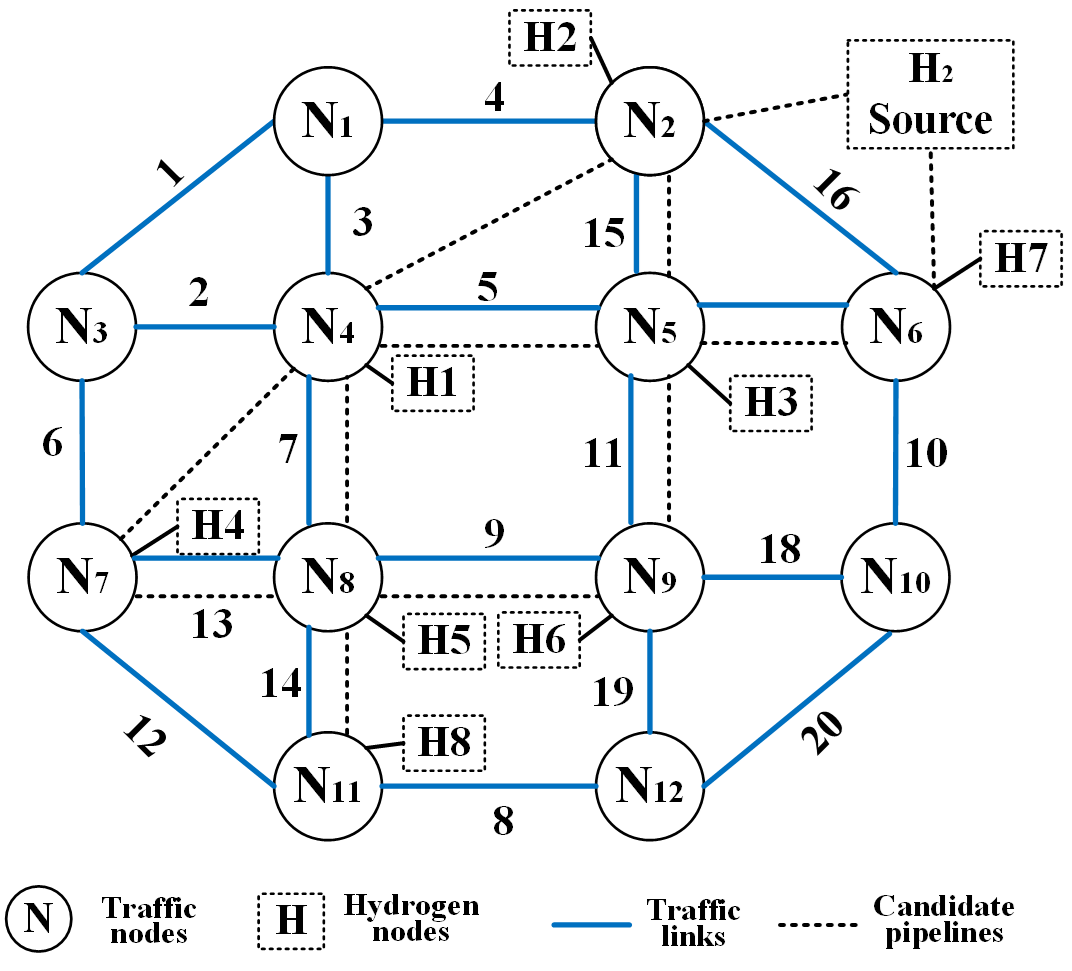}%
\label{fig_second_case}}
\caption{Network topologies: (a) power network; (b) transportation and hydrogen network.}\vspace{-0.3cm}
\label{fig_sim}
\end{figure}

\begin{figure}[htbp]
\centering
\subfloat[]{
\includegraphics[width=1.7in]{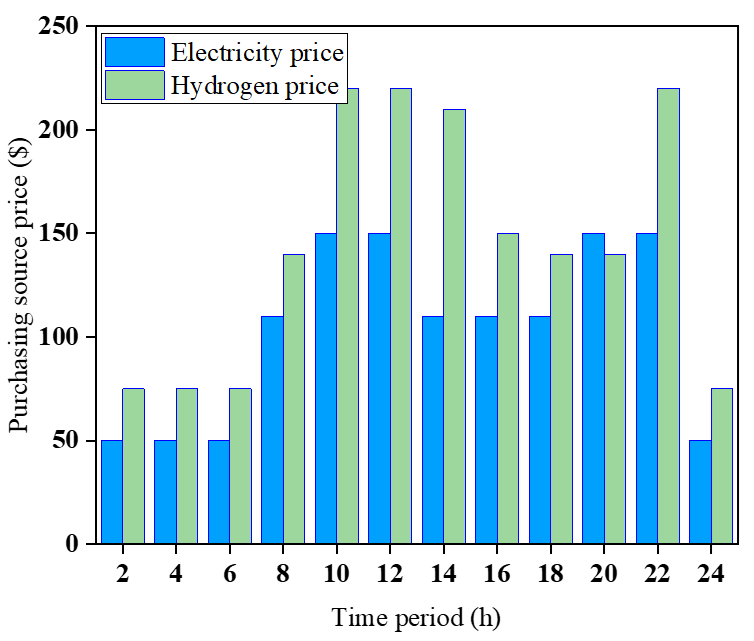}%
\label{fig_first_case}}
\subfloat[]{\includegraphics[width=1.7in]{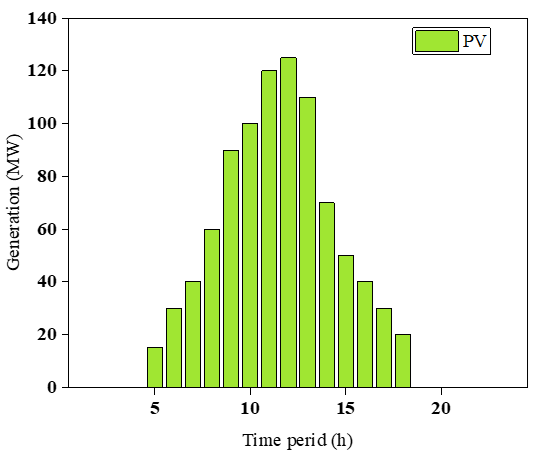}%
\label{fig_second_case}}
\caption{(a) Unit purchasing price of electricity and hydrogen from the main grid and hydrogen source; (b) Profile of PV generation forecast.}
\label{fig_sim}\vspace{-0.5cm}
\end{figure}

\vfill

\end{document}